\documentclass[3p,authoryear]{elsarticle}
\usepackage{graphicx}
\usepackage{adjustbox}
\usepackage{float}        
\usepackage{placeins}     
\usepackage{caption}      
\usepackage{setspace}     
\usepackage{caption}
\usepackage{subcaption}
\usepackage{amsmath}
\usepackage[colorlinks]{hyperref}
\usepackage[dvipsnames]{xcolor}
\usepackage[normalem,normalbf]{ulem}

\usepackage{pifont} 

\usepackage{graphicx}
\usepackage{tikz}
\usetikzlibrary{positioning}
\usepackage{pifont}
\usepackage{setspace}
\usepackage[fleqn]{nccmath} 
\usepackage{nicefrac}       
\usepackage{xcolor}         
\usepackage{graphicx} 
\usepackage{booktabs} 
\usepackage{siunitx}  
\usepackage{caption}  
\usepackage{array}    

\usepackage{amsmath}
\usepackage{mathrsfs} 
\usepackage{verbatim}

\usepackage{float} 
\usepackage{setspace}

\usepackage{comment}
\usepackage{tikz}
\usetikzlibrary{positioning}
\usepackage{pifont}

\usepackage{forest}

\usepackage{longtable}
\usepackage{multirow} 

\usepackage{hyperref}
\usepackage{lscape} 

\usepackage{soul}

\usepackage{url} 

\usepackage{booktabs}
\usepackage{tabularx} 

\usepackage{subcaption}
\usepackage{enumitem} 
\setlist[itemize]{align=parleft,left=12pt..2em} 
\setlist[enumerate]{align=parleft,left=12pt..2em} 
\newcommand{\note}[1]{\small{#1}} 

\usepackage{algorithm}
\usepackage{algpseudocode}
\usepackage{array}

\usepackage{bbm}

\begin{document}
\begin{titlepage}
    \centering
    \vspace*{2cm}

    {\LARGE\bfseries Profit-Aware Graph Framework for Cross-Platform Ride-Sharing: Analyzing Allocation Mechanisms and Efficiency Gains\par}

    \vspace{1.5cm}
    {\Large
    Xin Dong$^{a,*}$, Jose Ventura$^{b}$, Vikash V. Gayah$^{a}$
    \par}

    \vspace{1cm}
    {\small
    $^{a}$Department of Civil and Environmental Engineering, The Pennsylvania State University,\\
    University Park, PA 16802, United States \\
    Email: \texttt{xjd5036@psu.edu}, \texttt{gayah@engr.psu.edu}
    
    \vspace{0.5cm}

    $^{b}$Department of Industrial and Manufacturing Engineering, The Pennsylvania State University,\\
    University Park, PA 16802, United States \\
    Email: \texttt{jav1@psu.edu}
    }

    \vfill
    {\small
    $^{*}$Corresponding author.
    }
\end{titlepage} 
\title{Profit-Aware Graph Framework for Cross-Platform Ride-Sharing: Analyzing Allocation Mechanisms and Efficiency Gains}

\author[1]{Xin Dong\corref{cor1}}
\ead{xjd5036@psu.edu}
\author[1]{Jose Ventura}
\ead{jav1@psu.edu}
\author[1]{Vikash V. Gayah}
\ead{gayah@engr.psu.edu}
\cortext[cor1]{Corresponding author}
\address[1]{The Pennsylvania State University, University Park, PA, United States} 

\begin{abstract}
Ride-hailing platforms (e.g., Uber, Lyft) have transformed urban mobility by enabling ride-sharing, which holds considerable promise for reducing both travel costs and total vehicle miles traveled (VMT). However, the fragmentation of these platforms impedes system-wide efficiency by restricting ride-matching to intra-platform requests. Cross-platform collaboration could unlock substantial efficiency gains, but its realization hinges on fair and sustainable profit allocation mechanisms that can align the incentives of competing platforms.
This study introduces a graph-theoretic framework that embeds profit-aware constraints into network optimization, facilitating equitable and efficient cross-platform ride-sharing. 
Within this framework, we evaluate three allocation schemes---equal-profit-based, market-share-based, and Shapley-value-based---through large-scale simulations.
Results show that the Shapley-value-based mechanism consistently outperforms the alternatives across six key metrics. Notably, system efficiency and rider service quality improve with increasing demand, reflecting clear economies of scale. The observed economies of scale, along with their diminishing returns, can be understood with the structural evolution of rider-request graphs, where super-linear edge growth expands feasible matches and sub-linear degree scaling limits per-rider connectivity.
\end{abstract}


\maketitle

\section{Introduction}
Ride-sharing refers to the practice of using a single vehicle to serve multiple passengers simultaneously, each requesting distinct but directionally aligned trips, in an attempt to reduce the total number of vehicle trips and the associated driver labor required to meet travel demand \citep{shaheen2019shared, xu2022empirical, vignon2023regulating}. This concept has its roots in the carpooling initiatives that emerged during the US oil crisis of the 1970s \citep{ferguson1997rise}. Since then, ride-sharing has continuously evolved and been promoted in various forms, such as peer-to-peer ride-sharing \citep{tafreshian2020frontiers, zhong2020dynamic, dong2024analytical} and on-demand ride-sharing \citep{qin2022reinforcement, fielbaum2021demand}. The growing emphasis on ride-sharing is motivated by its wide-ranging benefits, including reductions in total vehicle travel distance, improvements in vehicle occupancy rates, lower travel costs, decreased fleet size requirements, mitigation of traffic congestion, and reductions in greenhouse gas emissions \citep{ke2023leveraging, pandey2025decline}. The sharing mobility economy has grown considerably in recent years and achieved more than \$130 billion in global consumer spending \citep{heineke2021shared} in 2019 (pre-pandemic). 
However, although existing ride-sharing systems have demonstrated considerable benefits, their full potential remains unrealized due to platform fragmentation, restricting matching opportunities and limit system-wide optimization efforts~\citep{sejourne2018price, chen2025scaling}. Empirical studies suggest that, in large metropolitan areas, a unified ride-sharing platform could increase share rates by up to 20\% and reduce vehicle miles traveled by approximately 0.25 miles per trip compared to the current fragmented landscape~\citep{liu2024impact}. These findings underscore the need for a deeper investigation into mechanisms that enhance the operational efficiency and structural performance of ride-sharing systems, particularly through cross-platform collaboration.

Addressing this need first requires understanding how rider matching has been optimized from an operational perspective. Extensive research has been devoted to optimizing rider matching within single-platform ride-sharing systems, employing heuristic algorithms~\citep{sundt2021heuristics}, combinatorial optimization techniques~\citep{masoud2017decomposition}, and graph-based formulations~\citep{meshkani2022generalized, alonso2017demand}. Among these, graph-based methods strike the most effective balance between scalability and solution quality, making them particularly advantageous for large-scale, real-time applications. However, these studies provide limited operational guidance on how to implement collaborative matching across competing platforms.

Beyond the operational challenge of enabling cross-platform collaboration, an equally important line of research has examined its potential system-level benefits. \citet{wang2022fed} demonstrate that platform collaboration can improve total revenue by 10–54\% compared to single-platform dispatching. \citet{GUO2023104397} explore various market structures aimed at dissolving segmentation in shared mobility markets, proposing designs that reduce total vehicle-miles traveled (VMT) by 6\%, serve 2.9\% more customers, and decrease total trips by 8.4\%. \citet{wang2023quantifying} show that integrating ride-sharing markets can enhance efficiency by 13.3\% (equivalent to 5\% fewer vehicle hours); however, they note that the resulting profit distribution tends to favor smaller platforms. Similarly, \citet{liu2024impact} find that a unified ride-sharing system can achieve share rates up to 20\% higher and reduce VMT by approximately 0.25 miles per trip compared to fragmented platforms. Collectively, these studies evaluate the system-level benefits of collaboration under the simplifying assumption that platforms are willing to cooperate rather than embedding it in incentive-aligned matching frameworks. Consequently, it remains unclear how these system-level benefits persist once collaboration is conditioned on incentive structures such as profit-sharing.

Another key open question concerns how ride-sharing systems scale and how cross-platform collaboration might reshape their structural properties. Prior research has revealed that ride-sharing systems exhibit strong scaling properties: the shareability---the feasibility of ride-sharing subject to time, distance, and capacity constraints---increases with demand density~\citep{santi2014quantifying}. \citet{tachet2017scaling} further demonstrate that shareability follows a universal scaling law: when trip data from different cities are rescaled by a dimensionless demand parameter specific to each city, they collapse onto a single curve. These findings echo the classic definition of economies of scale---efficiency improving as system size grows~\citep{stigler1958economies}---and, in ride-sharing systems, such gains are reflected in higher matching rates, shorter wait times, and reduced detours~\citep{lehe2021increasing}. Empirical evidence from different operational settings has further confirmed these effects. For example, \citet{liu2023scale} demonstrate that all three ride-hailing platforms in Chicago exhibited improved system performance as ride-pooling adoption scales. Using an entropy-based approach, \citet{liu2024impact} predict the potential performance gains of a unified ride-sharing system by quantifying how reduced platform competition levels can enhance overall efficiency. \citet{chen2025scaling} link system load with service rate and vehicle occupancy, and further generalize these scaling relationships across ten cities. While these studies underscore the scalability potential of ride-sharing systems, it remains unclear how such benefits evolve under collaborative regimes, particularly when shaped by different incentive mechanisms. Furthermore, while prior studies have primarily examined scaling effects in system performance, little is known about the structural evolution of ride-sharing networks. In particular, whether they follow patterns such as the Densification Power Law---where edges grow super-linearly with nodes, as established by \citet{leskovec2005graphs}---remains unclear.

As such, current research on ride-sharing collaboration faces three critical gaps: First, ride-sharing optimization studies remain confined to single-platform settings~\citep{ke2024emerging}, with limited exploration of multi-platform matching frameworks that incorporate profit allocation mechanisms into the matching process. Second, while scaling laws in competitive multi-platform ride-sharing systems are well-studied~\citep{liu2024impact}, the scaling behavior of profit-driven collaboration in multi-platform settings remains insufficiently understood. Third, despite the structural evolution of real-world social, technological, and information networks has been extensively studied~\citep{leskovec2005graphs}, how rider-request graphs evolve in ride-sharing systems---especially under profit-aware collaboration---remains unclear.
Our work addresses these gaps by conducting a comprehensive study of how profit-aware collaboration shapes both system efficiency and the structural evolution of rider-request networks in cross-platform ride-sharing. The primary contributions of this work include:
\begin{enumerate}
\item We propose a generalizable graph-theoretic model that integrates profit allocation constraints into the cross-platform matching process. Rider requests are modeled as a profit-weighted graph, with cross-platform edges activated only when profit splits ensure mutual benefit.

\item We conduct a comparative evaluation of three profit-sharing schemes---equal-profit, market-share, and Shapley-value-based---and assess their effects on six key performance indicators: share rate, vehicle miles traveled (VMT), platform profit, rider savings, detour distance, and wait time.

\item We demonstrate clear evidence of economies of scale in system performance: as demand increases, share rate rises and VMT per rider declines, with detour distances remaining stable and wait times decreasing under sufficient vehicle supply.

\item We present a novel perspective on the structural evolution of ride-sharing networks by uncovering super-linear edge growth and sub-linear degree scaling in rider-request graphs, illustrating how cross-platform collaboration bridges fragmented networks and enhances overall cohesion.
\end{enumerate}

The remainder of this paper is organized as follows. Section~\ref{sec: method} introduces our methodological framework, detailing the graph-based modeling approach and the profit-aware collaboration mechanisms. Section~\ref{sec: experiment} outlines the simulation setup, including the dataset, baseline experiment design, and sensitivity analysis procedures. Section~\ref{sec: system performance} analyzes system performance across six key metrics under various collaboration schemes. Section~\ref{sec: sensitivity} examines the robustness of these findings through sensitivity analyses. Finally, Section~\ref{sec: conclusion} summarizes the key findings, discusses the study's limitations, and suggests directions for future research. The Appendix further investigates the emergent structural scaling laws observed in the underlying graph.

\section{Methodology}
\label{sec: method}
This section presents the modeling framework for cross-platform ride-sharing. We begin by formulating pricing and feasibility criteria for rider pairs, establishing the conditions under which shared rides are both operationally viable and economically beneficial. Next, we introduce profit allocation mechanisms designed to assess the viability of inter-platform collaborations, ensuring that profit-sharing arrangements are mutually advantageous. Building upon these foundations, we construct the Rider–Rider Graph (RR-G), which encodes feasible and profitable rider pairings as weighted edges. This graph-based representation allows us to define and analyze three distinct collaboration modes: competition, full collaboration, and profit-aware collaboration. Finally, we employ a maximum-weighted matching algorithm to determine optimal ride-sharing pairs within the RR-G, maximizing overall system efficiency while adhering to the established constraints.

\subsection{Pricing and Profit Formulation}
A trip request from a given rider $r_i$ is defined by $r_i = \{P_i, t_i, O_i, D_i\} $, where $P_i$ is a categorical variable denoting the platform on which the trip is requested, $t_i$ is the request time, $O_i$ is the rider's origin, and $D_i$ is the rider's destination. We assume each  platform  offers an upfront  fare to this rider, \( f_i \), based on the direct trip distance from his origin to his destination, estimated travel time, and a base fare:

\begin{equation}
f_i = l_i (\alpha^d+\frac{\alpha^t}{\Bar{v}})(1-\beta) + \mu,
\label{eq: farei}
\end{equation}
where $l_i$ is the direct distance of $r_i$; $\alpha^d$, $\alpha^t$ are the unit costs associated with distance and time, respectively; $\Bar{v}$ is the average speed of the vehicles, which is assumed here to be a constant for simplicity;  $\mu$ denotes the fixed base fare, independent of travel distance or duration; and, $\beta$ represents the discount applied to the trip fare to incentive trip sharing.

We assume that payments to drivers are proportional to the vehicle's travel distance and time; let $o$ denote the driver payout rate. In a single-rider scenario, the platform collects a fare $f_i$ from rider $r_i$ and retains a fraction $(1 - o)$ as profit:

\begin{equation}
p_i = f_i (1 - o),
\label{eq: profiti}
\end{equation}

In a shared-trip scenario involving riders $r_i$ and $r_j$, the platform collects fares $f_i$ and $f_j$, and pays the driver based on the shared trip distance $l_{ij}$ and time-equivalent cost. The resulting platform profit is:

\begin{equation}
p_{ij} = f_i + f_j - l_{ij} \left( \alpha^d + \frac{\alpha^t}{\bar{v}} \right) o,
\label{eq: profitij}
\end{equation}

\noindent where the term $l_{ij} \left( \alpha^d + \frac{\alpha^t}{\bar{v}} \right) o$ represents the total driver compensation based on operational cost and payout rate.

\subsection{Feasible Pairs}
We define a pair of riders as a feasible pair if they can potentially share a trip. To identify such pairs, we propose a three-step framework that jointly enforces temporal, spatial, and profit-based feasibility. This approach ensures that shared trips are not only compatible in terms of timing and routing but also economically beneficial from the platform's perspective. Specifically, the framework evaluates:
\begin{itemize}
\item \textbf{Temporal feasibility}: Ensuring that the riders' requested pickup and drop-off times are sufficiently aligned to allow for a shared trip without significant delays.
\item \textbf{Spatial feasibility}: Assessing whether the riders' origins and destinations are geographically compatible, allowing for an efficient shared route.
\item \textbf{Profit-based feasibility}: Determining whether the shared trip would be financially advantageous for the platform, considering factors such as pricing strategies and operational costs.
\end{itemize}

\begin{figure}[htbp]
    \centering
    \includegraphics[width=0.7
    \linewidth]{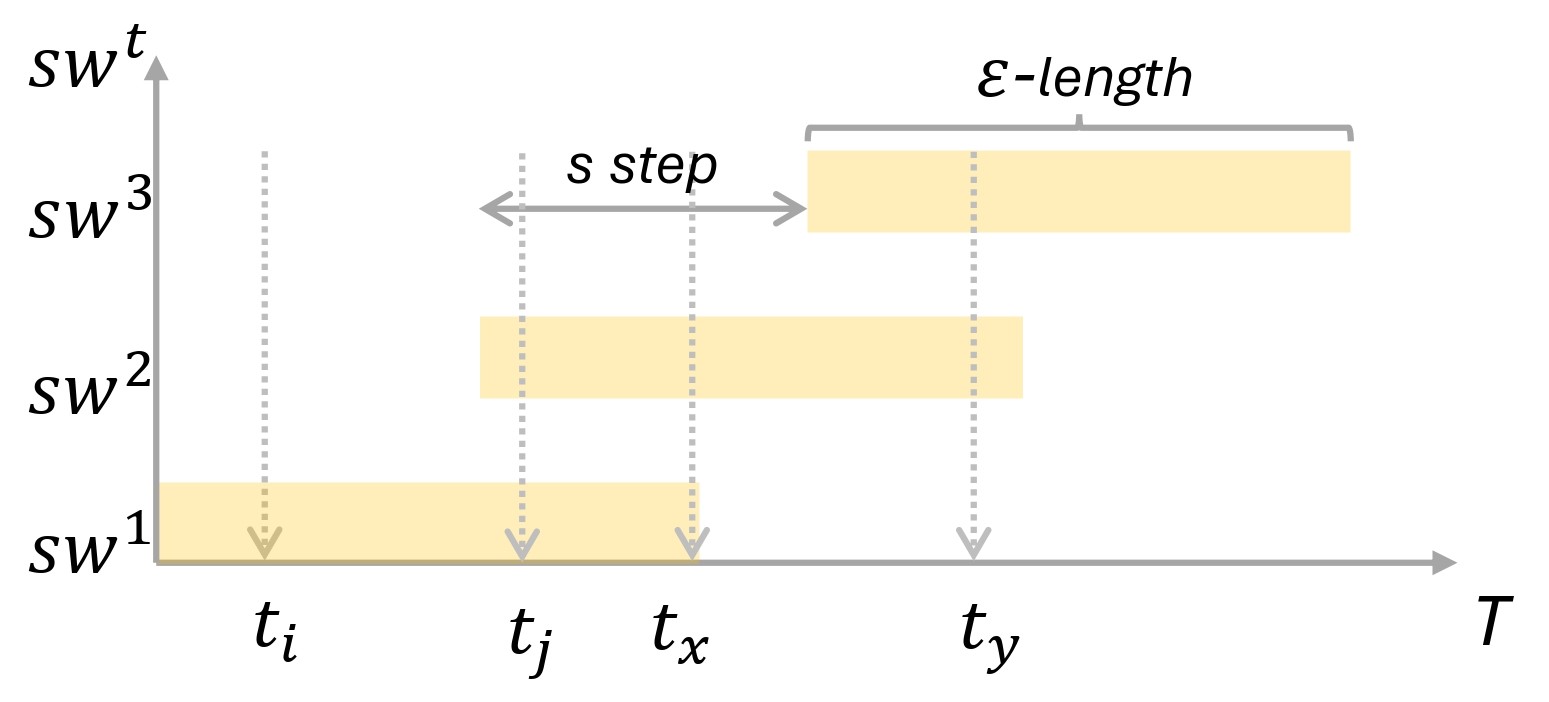}
    \newline\note{\small Note: \(t_i, t_j, t_x, t_y\) are rider arrival times, \(sw^t\) is the \(t^{th}\) sliding window, and \(\varepsilon, s\) are its length and moving step.}
    \caption{Illustration of a Sliding Window}
    \label{fig: sliding-window}
\end{figure}

First, to ensure temporal feasibility in ride-sharing, we employ a sliding window technique~\citep{datar2002maintaining} to segment trip requests into temporally localized batches. As illustrated in Figure~\ref{fig: sliding-window}, the sliding window, denoted as ${sw}^t$, is a temporal window of fixed length $\epsilon$ (e.g., 5 minutes) that moves forward in time with a step size $s$ (e.g., 1 minute). Each window defines a local time horizon over which rider requests are grouped for potential pairing. The step size $s$ determines the frequency at which new batches are formed; a smaller $s$ leads to more overlapping windows, allowing greater matching opportunities at increased computation cost. Riders whose request times fall within the current window $[t-\epsilon, t]$ are considered active. If a rider is not successfully matched within this window, they can still be considered during later windows, provided their maximum tolerable waiting time $\tau$ is not exceeded. This approach balances real-time responsiveness with the opportunity for efficient matching.

\begin{figure}[htbp]
    \centering
    \includegraphics[width=1\linewidth]{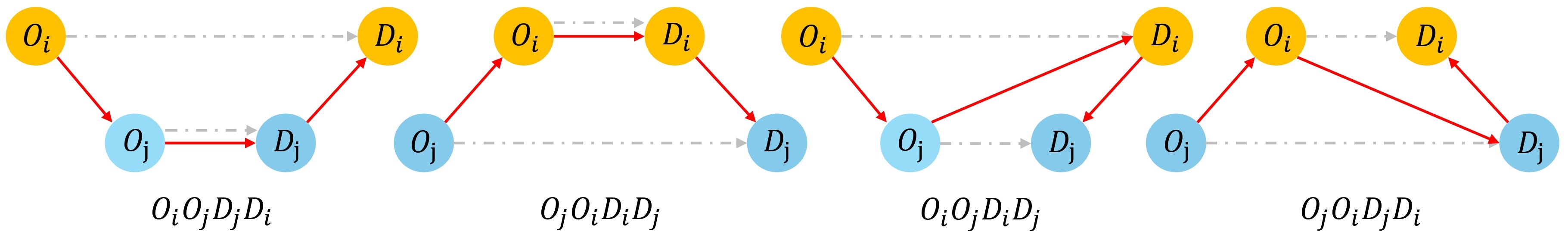}
    \centering{\small{\note{Note: The locations do not represent the actual locations but merely illustrate the path sequences. Dashed grey arrows show the direct single trip path, while the solid red arrows represent the shared trip path.}}}
    \caption{Illustration of Path Sequences}
    \label{fig: path}
\end{figure}

Second, we determine whether a spatially feasible path exists in which both riders may be served in a single vehicle. This involves evaluating the four possible sequences in which the two riders (e.g., $r_i$ and $r_j$) may be served: $O_i \underline{O_j D_j} D_i$, $O_j \underline{O_i D_i} D_j$, $O_i O_j D_i D_j$, and $O_j O_i D_j D_i$, as illustrated in Figure~\ref{fig: path}. In each sequence, the underlined segment represents a portion of the trip that would remain unchanged if the riders were served individually,  yielding no savings in travel distance. For each sequence, we compute the total travel distance by summing the shortest paths between consecutive locations along the route. These shortest paths are determined using Dijkstra’s algorithm \citep{dijkstra2022note} over the underlying road network graph. Among the four candidate sequences, the one with the minimum total distance is selected and denoted as the shared trip path with total travel distance represented by $l_{ij}$.


Finally, the selected shortest-distance path is validated through three operational constraints---wait time, detour ratio, and platform-level profit preservation---to ensure that the proposed sharing pair is not only feasible in space and time, but also economically viable for the platform: 

\begin{itemize}[leftmargin=*]
    \item Constraint 1 ($C_1$): \textbf{Waiting time constraints.} \\
    To reduce rider wait times, a maximum pickup delay $\tau$ is imposed in which each matched rider must be picked up within $\tau$ time units after their request time:
\end{itemize}

\begin{equation}
\frac{l_{V O_x}}{\Bar{v}} + t - t_x \le \tau, \quad \forall x \in \{i, j\}
\label{eq: c1}
\end{equation}

\hangindent=2em
\hangafter=0
\noindent where $t$ denotes the matching decision time, and $t_x$ is the request time of rider $x$ (\(x \in \{i, j\}\)). $l_{VO_x}$ represents the travel distance from the nearest vehicle to the pickup location of rider $x$ along the shared trip path defined above.

\begin{itemize}[leftmargin=*]
    \item Constraint 2 ($C_2$): \textbf{Detour distance constraints.} \\
    To prevent excessively long travel times, only a certain proportion of a passenger’s direct trip distance is acceptable as an additional distance as a part of a shared trip:
    \end{itemize}
    
\begin{equation}
\begin{aligned}
    & \quad \quad \frac{l_{O_xD_x} - l_x}{l_x} \le \gamma, \ \forall x \in {i,j},
\end{aligned}
\label{eq: c2}
\end{equation}
\hangindent=2em
\hangafter=0
\noindent where \(l_x\) denotes the direct trip distance of rider \(x\), and \(l_{O_xD_x}\) is the distance between the origin and destination of rider \(x\) (\(x \in \{i, j\}\)) in the shared trip, which may include partial detours. The parameter \(\gamma\) represents the maximum allowable detour, expressed as a fraction of the original trip distance.

\begin{itemize}[leftmargin=*]
    \item Constraint 3 ($C_3$): \textbf{Rideshare profit constraints.} \\
    The profit from the shared trips should be no less than the sum of profits if the trips were served individually \footnote {While some systems may allow this constraint to be relaxed as a soft condition, we treat it as a hard constraint in our framework due to the subsequent need for profit allocation among platforms.}:
\end{itemize}

\begin{equation}
\begin{aligned}
    & \quad \quad p_i + p_j \le p_{ij},
\end{aligned}
\label{eq: c3}
\end{equation}
\hangindent=2em
\hangafter=0
where $p_{ij}$ is from Equation~\eqref{eq: profitij}, and $p_i$ and $p_j$ are from Equation~\eqref{eq: profiti}.


\subsection{Profit Allocation Mechanisms}
\label{profit-aware}
Platforms must be fairly compensated for sharing their riders to enable sustainable cross-platform collaboration. Without a proper incentive structure, platforms may be reluctant to collaborate, especially when the gain from joint service is uncertain or uneven. To address this, three profit allocation mechanisms are considered in this paper: equal-profit scheme, market-share scheme, and  Shapley-value scheme.

For simplicity, we assume two platforms, labeled as $0$ and $1$, jointly participate in rider matching. The profit generated for each successfully matched rider pair is divided between the two platforms. 
Let $p(0)$ and $p(1)$ denote the standalone profit each platform would earn by serving its own riders independently, and let $p({0, 1})$ represent the profit when both platforms collaborate to serve the rider pair. The following subsections describe how the three allocation schemes split $p({0, 1})$ between platforms.

\subsubsection{Market-share Scheme}
The market-share scheme allocates the profit in proportion to each platform’s historical market share. Let $m_0$ and $m_1$ denote the normalized weights (e.g., based on past request volume or ride supply), with $m_0 + m_1 = 1$. These weights are computed over the entire historical period to reflect long-term participation levels. 
The profit allocation becomes:

\begin{equation}
\phi_0^{\text{market}} = m_0 \ p({0, 1}), \quad
\phi_1^{\text{market}} = m_1 \ p({0, 1})
\label{eq: market}
\end{equation}

\subsubsection{Equal-profit Scheme}
The equal-profit scheme simply divides the collaborative profit equally between platforms, ignoring their marginal contributions to the match. 

\begin{equation}
\label{eq: equal}
    \phi^{\text{equal}}_0 = \phi^{\text{equal}}_1 = \frac{p({0, 1})}{2}
\end{equation}

\subsubsection{Shapley-value Scheme}
To reflect the contribution of each platform in a fair and consistent manner, we adopt the Shapley-value-based profit allocation method~\citep{shapley1953value}. This method distributes the joint profit based on each platform’s marginal contribution across all possible platform coalitions.

Consider the more general case first i which \(N\)  participating platforms exist. For a given matched rider pair, the platforms that jointly serve it form a coalition \(S \subseteq N\), and the profit generated through this cooperation is denoted as \(p(S)\).  The Shapley value for platform \(i \in N\) is defined as:

\begin{equation}
\label{eq: shapley}
\phi_i^{\text{Shapley}} = \sum_{S \subseteq N \setminus \{i\}} \frac{|S|! \cdot (|N| - |S| - 1)!}{|N|!} \left[p(S \cup \{i\}) - p(S)\right]
\end{equation}

\noindent where $\phi_i$ is the profit allocated to platform $i$; \(p(S \cup {i})\) is the value of the coalition with the addition of player \(i\); summation over \( S \subseteq N \setminus \{i\} \) considers all coalitions without player \(i\). \(|N|\) is the number of players, \(|S|\) is the number of players in a formed coalition. 

For the two-platform setting considered in this study, where \(N = \{0, 1\}\), we define the additional profit generated by serving a rider pair jointly rather than separately as:

\begin{equation}
\Delta p = p({0, 1}) - p(0) - p(1)
\label{eq: dp}
\end{equation}

\noindent
Using this, the Shapley values simplify to:

\begin{equation}
\phi_0^{\text{Shapley}} = p(0) + \frac{1}{2} \Delta p, \quad
\phi_1^{\text{Shapley}} = p(1) + \frac{1}{2} \Delta p
\label{eq: shapley-simple}
\end{equation}

This formulation guarantees that each platform receives at least its standalone profit, with the additional gains from collaboration split evenly. Compared to the equal-profit and market-share schemes, the Shapley-value approach explicitly accounts for marginal contributions and provides a structured and equitable method for allocating collaborative profits.


These allocation rules serve as the basis for evaluating the financial viability of inter-platform cooperation, which will guide collaboration decisions in the profit-aware matching framework discussed next.

\subsection{Rider-Rider Graph}
We construct a series of undirected, weighted graphs—--referred to as Rider–Rider Graphs (RR-G)—--to capture the adjacency relationships among riders over discrete time intervals. In this framework, each node represents an individual rider request, and edges denote feasible and mutually beneficial pairing opportunities, weighted by the associated profit potential. This graph-based approach enables flexible modeling of both collaborative and competitive multi-platform interactions, serving as the foundation for identifying feasible and profitable rider matches under various operational scenarios.

At time $t$, the RR-G is denoted by \( RR-G^t = (V^t, E^t) \), where \( V^t \) is the set of active riders in the $t^\text{th}$ sliding window \( sw^t \), and \( E^t \) is the set of edges representing feasible pairwise connections among them. The number of vertices, \( |V^t| \), corresponds to the number of active riders in the window, and the edge set is defined as \( E^t \subseteq \{(x, y) \mid x, y \in V^t, \, x \neq y\} \).

To determine whether an edge should exist between any two riders, we construct a binary indicator matrix \( I^t \), which encodes the feasibility of each rider pair based on the constraints defined in Equations~\eqref{eq: c1}--\eqref{eq: c3}:

\begin{equation}
\begin{aligned}
    I^t &= \left\{ I_{ij}^t \mid \forall i, j \text{ such that } r_i, r_j \in \text{sw}^t \right\}, 
    \quad \text{where} \quad
    I_{ij}^t =
    \begin{cases} 
    1 & \text{if } C_1 \cap C_2 \cap C_3, \\
    0 & \text{otherwise}.
    \end{cases}
\end{aligned}
\label{eq: I}
\end{equation} 
\noindent where \(i\) and \(j\) are indices of riders in the $t^\text{th}$ sliding window, and \(C_1\), \(C_2\), and \(C_3\) refer to the feasibility constraints.

Each edge in the RR-G is associated with a dynamic weight that reflects the quality of the connection between two riders. The edge weight, defined in Equation~\eqref{eq: wij}, consists of two components: the profit of the shared trip and the sum of waiting times for the two connected riders. 

\begin{equation}
w_{ij} = I_{ij}^t \left[ p_{ij} + \psi(t_i^w + t_j^w)\,\alpha^t \right],
\label{eq: wij}
\end{equation}

\noindent where \( t_i^w \) and \( t_j^w \) denote the waiting times of riders \( r_i \) and \( r_j \), respectively; \( p_{ij} \) is the profit of the shared trip; \( \psi \in [0, 1] \) is a weight factor balancing profit and waiting time; and \( \alpha^t \) is a scaling parameter. The indicator \( I_{ij}^t \), computed from Equation~\eqref{eq: I}, ensures that only feasible rider pairs are considered in the graph. 

\begin{figure}[H]
    \centering
    \small
    \begin{minipage}[t]{0.53\linewidth}
        \centering
        \begin{tabular}{cccccc}
        \toprule
        RID & Platform & Time & OID & DID & Edge Weight\\
        \midrule
        1 & 0 & 0  & 1 & 5 & 6.55\\
        2 & 0 & 1 & 3 & 1 & 3.17\\
        3 & 0 & 2 & 4 & 5 & 5.67\\
        4 & 0 & 3 & 5 & 2 & 3.21\\
        5 & 1 & 0 & 1 & 1 & 0.19\\
        6 & 1 & 2 & 1 & 1 & 0.33\\
        7 & 1 & 3 & 1 & 6 & 3.01\\
        8 & 1 & 3 & 3 & 3 & 0.31\\
        9 & 1 & 4 & 5 & 2 & 3.37 \\
        10 & 1 & 5 & 5 & 6 & 4.09\\
        \bottomrule
        \end{tabular}
    \end{minipage}%
    \begin{minipage}[T]{0.47\linewidth}
        \centering
        \includegraphics[height = \linewidth]{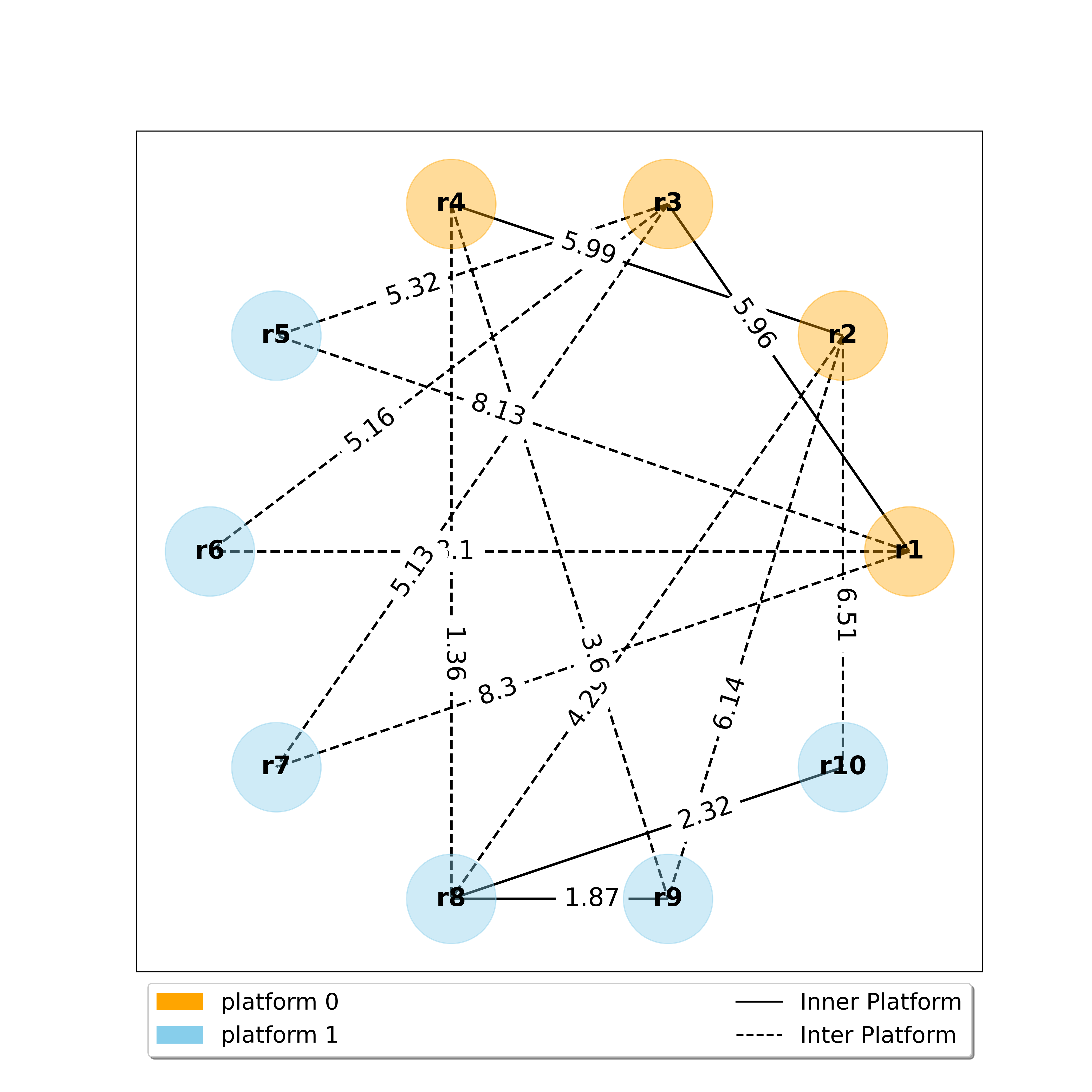}  
    \end{minipage}
    \small{\note{Note: Solid lines show feasible pairs within platforms, and dashed lines show those between platforms.}}
    \caption{Rider Requests and Schematic Rider-Rider Graph at a single time step}
    \label{fig: sub_rr_graph}
\end{figure}

Figure~\ref{fig: sub_rr_graph} illustrates an example of $RR$–$G^t$ for a single time window. 
The left table summarizes the rider request information in the current sliding window $[0, 5]$, including their platform affiliation and weights of feasible pairings. The right graph visualizes the constructed RR–$G^t$ based on this data, where edge weights correspond to the computed values in Equation~\eqref{eq: wij}. Note that different colors are assigned to the vertices, representing the riders belonging to the two different platforms.
    
The generic graph $RR-G^t$ is then adapted in the following subsections to reflect different scenarios: competition (in which each platform operates independently), full collaboration (in which platforms share all trip request information) and profit-aware collaboration (in which platforms only share trip request information with the other platform when advantageous to do so). 

\subsubsection{Competition} 
In a competitive scenario, each platform operates independently without sharing rider information. 
Given the RR-G of the entire ride-sharing system during the \( t^{\text{th}} \) sliding window, each platform \(i\) maintains an induced subgraph \( H_i^t \) that consists of vertices from \( RR-G^t \) and all the edges from \( RR-G^t \) that connect its own riders. If there are \( k \) different platforms, there will be \( k \) induced sub-graphs as defined in Equation~\eqref{eq: H}. The weight of edges for vertices in the same sub-graph is calculated using Equation~\eqref{eq: wij}, and any \(r_i\) and \(r_j\) that belong to different platforms will not interact in any way.

\begin{equation}
    H^t_i = (V_{H_i}^t, E_{H_i}^t), \text{ where } V_{H_i}^t = \{x \mid x \in P_i\} \text{ and } E_{H_i}^t = \{(x,y) \mid x, y \in P_i\},
    \label{eq: H}
\end{equation}
where \(P_i\) represents the \(i^{th}\) platform (\(i \in [1,k]\)), \(V_{H_i}^t\) is the vertex set with riders belonging to the \(i^{th}\) platform (\(V_{H_i}^t \subseteq V^t\)), and \(E_{H_i}^t\) is the edge set connecting only riders within the platform (\(E_{H_i}^t \subseteq E^t\)). Vertices $x,y$ are riders in $P_i$.
 
\subsubsection{Full Collaboration}
Full collaboration assumes the individual platforms share all rider request information with each other. In this case, the individual platforms operate as one; thus, the collaborative graph \(C^t = (V_C^t, E_C^t)\) includes all vertices and edges from \( RR-G^t \) with weights defined by Equation~\eqref{eq: wij}:
\begin{equation}
    C^t = (V_C^t, E_C^t) \text{, where } V_C^t = V^t \ \text{ and } \ E_C^t = E^t, 
    \label{eq: C}
\end{equation}
\noindent where $V_{C}^t$ is the vertex set with active riders in the current sliding window, and $E_{C}^t$ is the edge set with edges connecting both inter-platform riders and intra-platform riders.

\subsubsection{Profit-aware Collaboration}

Profit-aware collaboration introduces platform-level economic constraints to determine whether a cross-platform edge should remain active. Specifically, each platform evaluates the projected profit of sharing against two benchmarks: the profit from serving the rider as a single trip ($p_i$ or $p_j$), and the intra-platform shared-trip profit ($\phi^{\text{inner}}$), both computed using the same allocation mechanism. 

To incorporate profit considerations into the matching process, we define a profit-aware feasibility matrix $\tilde{I}{ij}^t$, which refines the original feasibility matrix $\bar{I}{ij}^t$ (Equation~\eqref{eq: I}). The matrix $\bar{I}{ij}^t$ captures whether the rider pair $(i, j)$ satisfies basic matching constraints such as time-window overlap and detour limits. We further filter these feasible pairs by requiring that both platforms gain no less profit from collaboration than from serving their riders independently:

\begin{equation}
\tilde{I}_{ij}^t = \mathbbm{1} \left(
(\phi_{r_i}^{\text{inter}} \ge \phi_{r_i}^{\text{inner}} )\land
(\phi_{r_i}^{\text{inter}} \ge p_i) \land
(\phi_{r_j}^{\text{inter}} \ge \phi_{r_j}^{\text{inner}}) \land
(\phi_{r_j}^{\text{inter}} \ge p_j)
\right)
\cdot \bar{I}_{ij}^t
\label{eq: Ituda}
\end{equation}

\noindent where $\mathbbm{1}(\cdot)$ is an indicator mask that returns 1 if all profit conditions are met. $\phi^{\text{inter}}$ and $\phi^{\text{inner}}$ denote the expected platform profits under inter- and intra-platform sharing, respectively, using a consistent allocation mechanism. 

Then, the edge weights are updated with the profit-aware feasibility matrix $\tilde{I}_{ij}^t$ and Equation~\eqref{eq: wij}. The resulting profit-aware collaboration graph, denoted as $RC^t = (V_{RC}^t, E_{RC}^t)$, is constructed by retaining only those edges that satisfy both feasibility and profit-aware conditions.

\begin{equation}
    RC^t = (V_{RC}^t , E_{RC}^t ), \
    \text{where} \
    V_{RC}^t = V^t \text{ and }  \
    E_{RC}^t = \{(i,j) \in E^t \mid \tilde{I}_{ij}^t = 1\}
    \label{eq: RC}
\end{equation}

\noindent where $V_{RC}^t$ includes all active riders in the current window (i.e., all vertices in $V^t$), while $E_{RC}^t$ consists of intra-platform edges and those inter-platform edges that satisfy both feasibility and profit-aware conditions, as encoded in $\tilde{I}_{ij}^t$.

\subsection{Maximum-Weighted Matching (MWM)}
This section describes the MWM method used to determine rider pairs to form shared trips. Let $G = (V, E)$ be an undirected weighted general graph, which can represent any of the previously described RR-G graphs above, $H^t = (V_{H}^t, E_{H}^t)$, $C^t = (V_{C}^t, E_{C}^t)$, and $RC^t = (V_{RC}^t, E_{RC}^t)$.
A potential match is a subset of edges $M \subseteq E$ such that each node in $G$ has at most one incident edge in the subset $M$. The optimum match obtained with the MWM algorithm, $M^*$, is the subset of edges that yield the  maximum sum of weights of selected edges. This optimum match would need to be provided  dynamically over each sliding window under different market scenarios, and would be obtained via the following optimization:

\begin{equation}
\begin{aligned}
    \max \quad & W(G) = \sum_{E_{ij} \in M} w(E_{ij}) \\
    \text{s.t.} \quad & E_{ij} \cap E_{xy} = \emptyset, \quad \forall E_{ij}, E_{xy} \in M, \\
    & M \subseteq E, \\
    & w(E_{ij}) = w_{ij},
\end{aligned}
\label{eq: mwm}
\end{equation}
where $G$ is one of the defined RR-Gs, $W(G)$ is the maximum weight of graph $G$, and $E$ is the corresponding edge set of the RR-G. $M$ is the matching set to be assigned, which is a subset of $E$.
 
Edmonds~\citep{edmonds1965maximum} provided the first polynomial-time algorithm for solving MWM in general graphs, based on augmenting paths that alternate between matched and unmatched edges. For sparse graphs such as our rider-rider graphs, we adopt the $O(mn\log n)$ algorithm by Galil~\citep{galil1986efficient}, which optimizes Edmonds’ Blossom algorithm with dynamic trees and is well-suited for our sparse rider-rider graphs. 

\begin{figure}[H]
    \centering
    \includegraphics[width=1\linewidth]{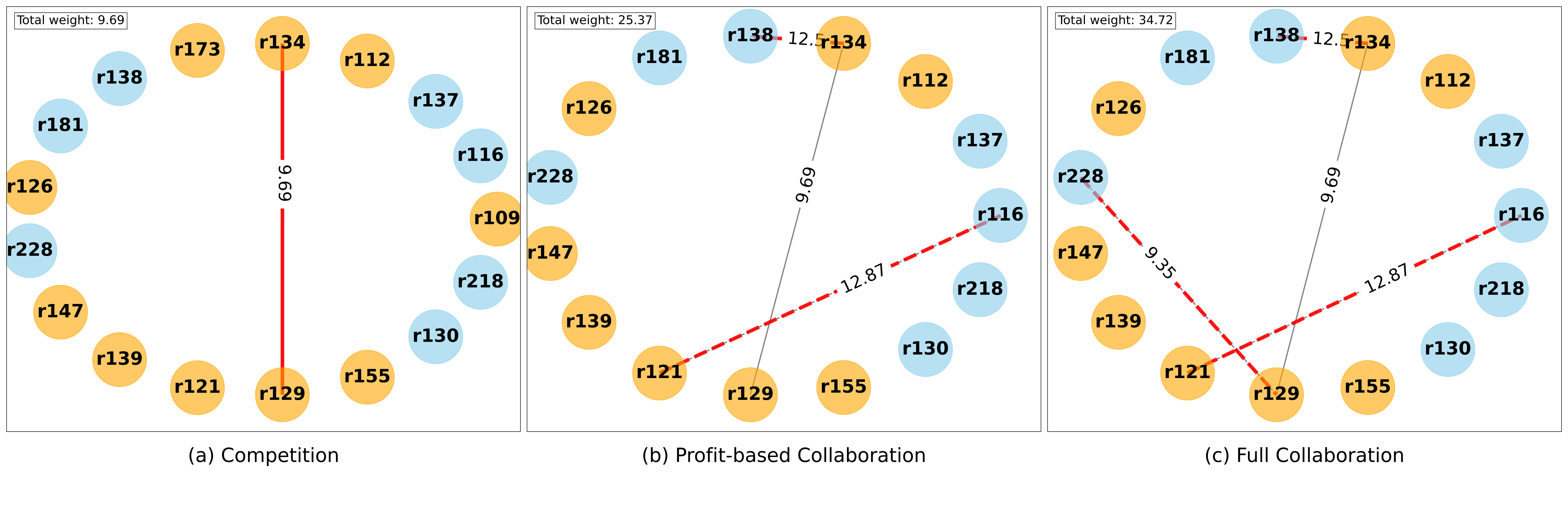}
    \caption{Maximum-Weighted Matching results within a single sliding window for (a) Competitive Market, (b) Profit-aware Collaboration, and (c) Full Collaboration. Yellow and blue nodes represent rider requests from two different platforms. Solid and dashed lines show intra- and inter-platform feasible pairs. Red lines indicate the selected matches within the current window, with the total matching weight shown in each subfigure.}
    \label{fig:match_example}
\end{figure}

To illustrate how the MWM algorithm operates under different collaboration modes, Figure~\ref{fig:match_example} presents an example result within a 5-minute sliding window. Each node represents a rider's trip request, with yellow and blue indicating requests from two ride-sharing platforms. Solid black lines denote potential intra-platform matches, and dashed lines represent inter-platform matches. 
Figure~\ref{fig:match_example}(c) has fewer dashed lines due to the stricter constraints of profit-based collaboration. In the competition scenario, only solid lines are eligible for matching, while both solid and dashed lines can be used under the collaboration scenarios (Figures~\ref{fig:match_example}(b) and (c)). The selected optimal matches are shown in red, with total matching weight displayed above each sub-figure.

\section{Experiments}
\label{sec: experiment}
This section presents simulation experiments designed to evaluate the proposed framework under the different market scenarios described above: competition, full collaboration, and profit-aware collaboration (under three different mechanisms). To capture the impact of collaboration modes on ride-sharing performance, we simulate profit allocation for each shared trip and assess how it affects system-level outcomes. The following subsections detail the data used to support the simulation, parameters, experimental procedures, and evaluation metrics.

\subsection{Data}
\begin{figure}[htbp]
    \centering
    \includegraphics[width= 1\linewidth]{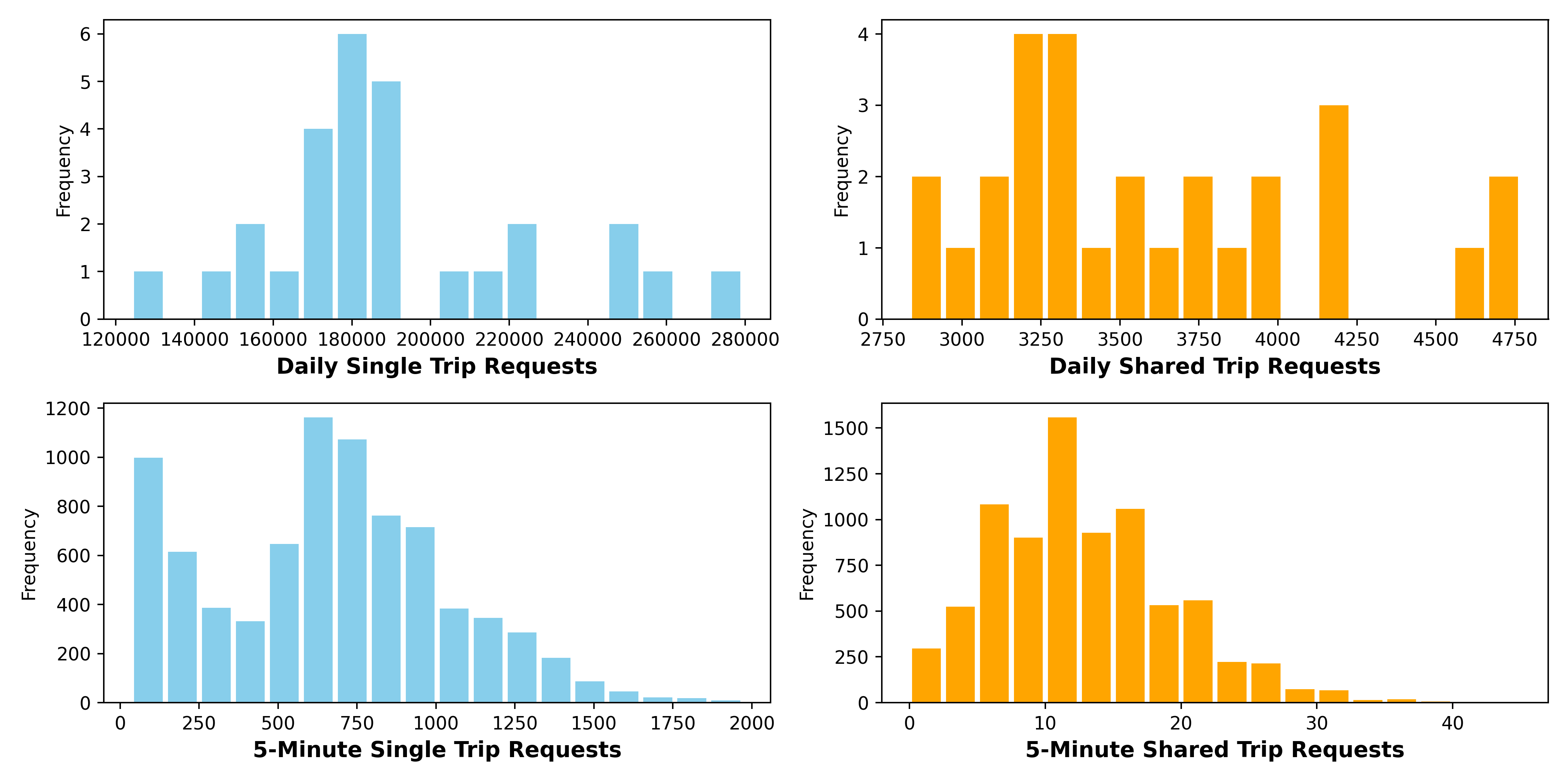}
    \caption{Distribution of Aggregated (Every 5-minute) Trips in February, 2023}
    \label{fig: trips summary}
\end{figure}

TLC trip data \citep{tlc} from Manhattan, NYC are used as system inputs. This includes pickup and drop-off locations, timestamps, and other relevant features. The dataset is publicly available and has been widely adopted in transportation studies \citep{ying2021auto, chen2023quantifying, shulika2024spatiotemporal}.
Figure~\ref{fig: trips summary} presents the daily and five-minute aggregated counts of single and shared trip requests. In February 2023, the system recorded approximately 180,000 single-trip and 3,250 shared-trip requests per day, corresponding to about 750 and 10 requests per 5-minute interval, respectively. These values highlight the relatively low level of ride-sharing in the system, possibly due to the limited number of shared trip requests. Inter-platform collaboration could increase the likelihood of finding compatible pairs by expanding the pool of requests that may be matched. To represent a high-demand real-world scenario that creates favorable conditions for testing collaborative mechanisms, we select February 4th, 2023---the day with the highest recorded number of shared trip requests (4,235)---as the simulation date.

To simulate travel distances more precisely, we use the GeoPandas \citep{jordahl2014geopandas} and OSMNX \citep{boeing2017osmnx} Python packages to map trip origins and destinations---originally encoded in 63 taxi zones in Manhattan---to 4,589 intersections in the roadway network. A distance matrix is precomputed to estimate shortest-path travel distances between all node pairs. Each trip is then assigned an origin and destination node within its respective taxi zone to best replicate the reported trip length. This spatial decoding supports accurate estimation of pairwise rider distances and enables graph construction with realistic edge weights.

\subsection{Baseline Experiment Setup}

To evaluate the impact of different collaboration scenarios, we design a baseline experiment for a ride-sharing market with two available platforms (marked as 0 and 1). The collaboration setting varies across competition, full collaboration (unrestricted), and collaboration under three profit-aware schemes, with all other system parameters held constant.

Table~\ref{tab:parameter-setting} summarizes the key simulation settings. The simulation runs over a 24-hour horizon ($H = 1440$ minutes) using a 5-minute sliding window. We assume a fixed platform commission rate ($o = 0.5$), average vehicle speed ($\bar{v} = 15$ mph), and a fare structure with $\alpha^d = 2$, $\alpha^t = 0.5$, and base fare $\mu = 2$\$. For the baseline configuration, three key behavioral parameters are fare discount factor $\beta = 0.1$, detour upperbound $\gamma = 0.2$, and rider waiting time threshold $\tau = 10$ minutes. Sensitivity analysis on these parameters is discussed in Section~\ref{sec: sensitivity}.

\begin{table}[htbp]
\centering
\caption{Summary of parameters in ride-sharing simulation}
\begin{tabular}{@{}cll@{}}
\toprule
Notation  & Interpretation & Value  \\ 
\midrule
$n$ & Number of replications & 5 \\
$H$ & Horizon (minute) & 1440 \\ 
$t$ & Time step & $[0,H]$ \\
$M$ & The number of Platforms & 2 \\ 
$Q^t$ & Riders' trip request queue at certain time & ${r_1, r_2, ..., r_i}$, $i\in[1, 4235]$ \\
$\epsilon$ & Length of sliding window (minute) & 5 \\ 
$s$ & Moving step of sliding window (minute) & 5 \\ 
${sw}^t$ & Sliding window (minute)  & $[t-\epsilon,t]$ \\
$P$ & Market Share & {0.58, 0.42} \\ 
$\Theta$ & Platform collaboration willingness & 1.0 \\ 
$\tau$ & Maximum waiting time (minute) & {5, 10, 15, 20} \\ 
$o$ & Platform commission rate & 0.5 \\ 
$\gamma$ & Maximum detour factor & {0.2, 0.4, 0.6, 0.8} \\ 
$\Bar{v}$ & Velocity (mph) & 15 \\ 
$\alpha^d$ & Value of distance (vod) (\$/mile) & 2 \\ 
$\alpha^t$ & Value of time (vot) (\$/minute) & 0.5 \\
$\mu$ & Fixed base fare & 2 \\
$\beta$ & Discount ratio of trip fare & {0, 0.1, 0.2, 0.3} \\
$\psi$ & Weight of waiting time considered in edge calculation & [0,1]\\
\bottomrule
\end{tabular}
\label{tab:parameter-setting}
\end{table}

\subsection{Simulation Procedure} 

We simulate ride-sharing operations over a 24-hour horizon using a discrete-time framework with a 5-minute sliding window. All trip requests are preloaded into the simulator, characterized by their arrival time, platform, origin, and destination. Calculated trip fares. A trip becomes active only when it enters the current sliding window $sw^t$. At each time step $t$, the simulator: (i) updates rider queues; (ii) computes the shareability matrix $I^t$; (iii) constructs three rider–rider graphs under different collaboration settings—competitive ($H^t$), full collaboration ($C^t$), and profit-aware ($RC^t$); (iv) evaluates projected profits using predefined allocation mechanisms; and (v) solves MWM to assign shared trips. The full simulation workflow is outlined in Algorithm~\ref{alg: simulation}.

\begin{algorithm}
\caption{Simulation Procedure over One Day ($H = 1440$ minutes)}
\label{alg: simulation}
\begin{algorithmic}[1]
\State \textbf{Initialize} simulation:
    \State \hspace{1em} Set time horizon $H = 1440$ minutes (24 hours)
    \State \hspace{1em} Set sliding window size $\epsilon = 5$ minutes and step size $s = 5$
    \State \hspace{1em} Load trip queue $Q^0$ with all requests (arrival time, platform, origin, destination)
    \State \hspace{1em} Reassign platforms randomly according to market share $m$, if re-selection is required
    \State \hspace{1em} Compute fares using Equation~\eqref{eq: farei} with $\alpha^d = 2$, $\alpha^t = 0.5$, and base fare $\mu = 2\text{\$}$

\For{$t = 0$ \textbf{to} $H - \epsilon$ \textbf{step} $s$}
    \Procedure{SimulateStep}{$t$}
        \State \textbf{Update} rider queue $Q^t$, including rider statuses (e.g., arrived, waiting, served) and waiting times
        \State \textbf{Compute} shareability indicator matrix $I^t$ using Equations~\eqref{eq: c1}--\eqref{eq: I}
        \State \textbf{Construct} rider–rider graphs $RR$-$G^t$:
            \State \hspace{1em} $H^t \gets$ competitive graph (intra-platform only; Eq.~\ref{eq: H})
            \State \hspace{1em} $C^t \gets$ full collaboration graph (all feasible pairs; Eq.~\ref{eq: C})
            \State \hspace{1em} $RC^t \gets$ profit-aware graph using $\tilde{I}_{ij}^t$ (Eq.~\ref{eq: Ituda},~\ref{eq: RC}) with profit-aware allocation rules (Eq.~\eqref{eq: market}--\eqref{eq: shapley})
        \For{each graph $G^t \in \{H^t, C^t, RC^t\}$}
            \State \textbf{Solve} maximum-weighted matching $M^* \gets \textsc{MWM}(G^t)$ (Eq.~\eqref{eq: mwm})
            \State \textbf{Update} $Q^t$ for shared trips and assign profits to platforms accordingly
        \EndFor
    \EndProcedure
\EndFor
\end{algorithmic}
\end{algorithm}

\subsection{Evaluation Metrics}
To compare performance under different collaboration settings, we evaluate six key metrics that quantify system performance from both the platform and rider perspectives:

\begin{itemize}
    \item \textbf{Share rate:} percentage of rider requests served as shared trips.
    \item \textbf{Vehicle miles traveled (VMT):} distance traveled by vehicles to serve riders. 
    \item \textbf{Platform profit:} total revenue generated by the trip fare minus the operational costs for each platform.
    \item \textbf{Average rider savings:} the average rider's savings over the shared trip compared with the corresponding single trip.
    \item \textbf{Passenger detour distance:} additional distance a passenger has to travel due to sharing a ride with others, compared to a direct trip.
    \item \textbf{Passenger waiting time:} time a rider waits from requesting a ride until being offered a trip.
\end{itemize}

\newpage
\section{System Performance}
\label{sec: system performance}


\subsection{Share Rate} 
\begin{figure}[htbp]
    \centering
    \includegraphics[width=1\linewidth]{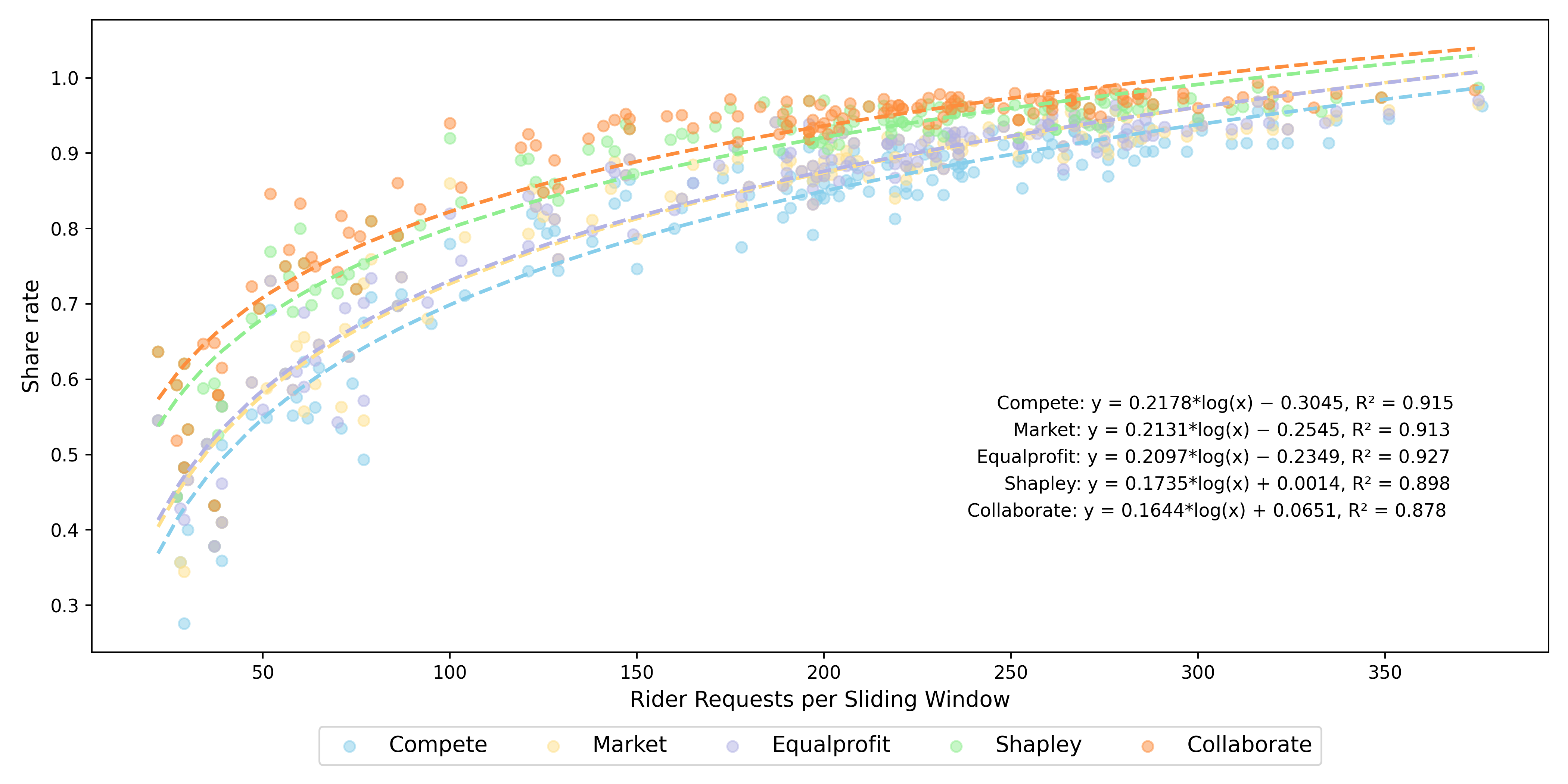}
    \caption{Performance Scaling: Sharing Rate vs. Rider Requests per Sliding Window. All scenarios—competition, profit-aware collaboration (market share-based, equal profit-based, and Shapley-based mechanisms), and full collaboration—follow a sub-linear scaling with a consistently strong fit.}
    \label{fig: share_rate_scaling}
\end{figure} 

\begin{figure}[H]
    \centering
    \includegraphics[width=1\linewidth]{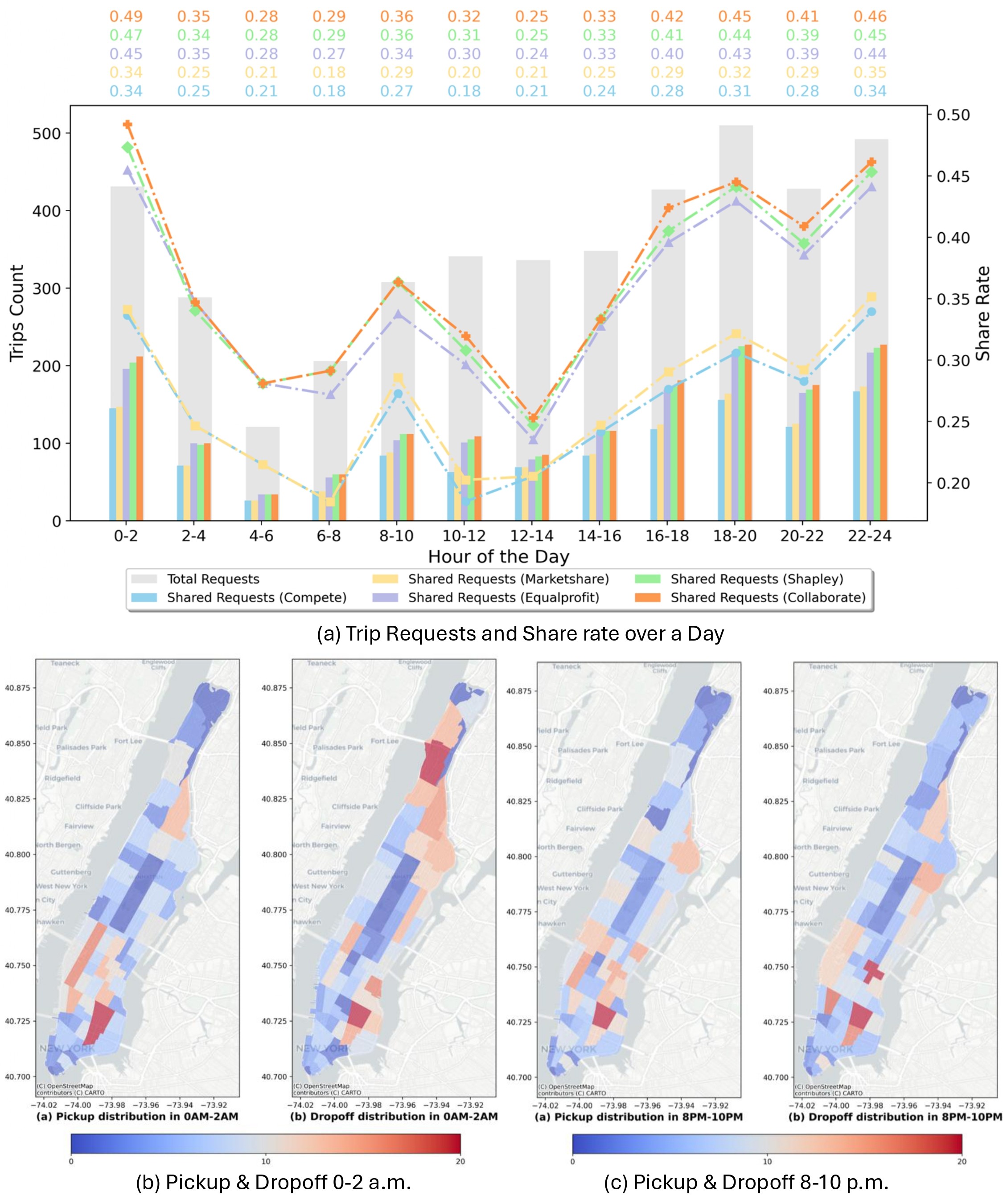}
    \caption{Share Rate and Trip Distribution. (a) Shared Rate in two-hour intervals over One Day; (b) Spatial distribution of trip requests between 0 AM and 2 AM; (c) Spatial distribution of trip requests between 8 PM and 10 PM. The grey bars show the total number of shared trip requests during each interval, the colored bars present the number of realized shared trips under different scenarios, and the share rates are plotted as dashed lines. }
    \label{fig: spatial_temporal}
\end{figure}

Figure~\ref{fig: share_rate_scaling} shows that the share rate consistently increases  with the number of rider requests across all scenarios, with full collaboration and Shapley-value-based collaboration yielding notably higher rates than other scenarios. All scenarios exhibit the most rapid growth in share rate during the early stages of system expansion, reflecting pronounced initial scale effects. As rider volume increases, share rates continue to rise across all scenarios, albeit at a decreasing rate. This flattening trend in the upper range suggests diminishing marginal returns, as the system approaches saturation in its ability to form additional successful matches. Such saturation may stem not only from aggregate system size but also from structural and operational constraints inherent in ride-sharing systems--—including spatial-temporal dispersion of requests, limitations in fleet routing, and restrictions imposed by the matching mechanisms. One illustrative case is shown in Figure~\ref{fig: spatial_temporal}(a), where time intervals with similar overall demand---such as 0-2 a.m. and 8-10 p.m.---exhibit notably different share rates (0.49 vs. 0.41).
Figure~\ref{fig: spatial_temporal}(b) and (c) further illustrate this contrast, showing denser clustering of pickup and drop-off locations during late-night hours. This spatial concentration likely facilitates route overlap and increases shareability, whereas the more dispersed pattern observed in the evening reduces matching opportunities.

To better quantify this relationship and test for statistical significance, we fit a log-linear regression model of the form $\text{Share Rate} = \beta \log(N) + \alpha$, where $N$ is the number of rider requests every 10 minutes. 
All mechanisms exhibit strong and persistent improvements in share rate as the system scales, with estimated $\beta$ values ranging from 0.1644 (Full Collaboration) to 0.2178 (Competition), and notably high $R^2$ values ranging from 0.878 (Full Collaboration) to 0.927 (Equal-profit-based Collaboration). (All coefficients are statistically significant with 99.9\% confidence.)

 

\subsection{Vehicle Miles Traveled}
\begin{figure}[H]
    \centering
    \includegraphics[width=1\linewidth]{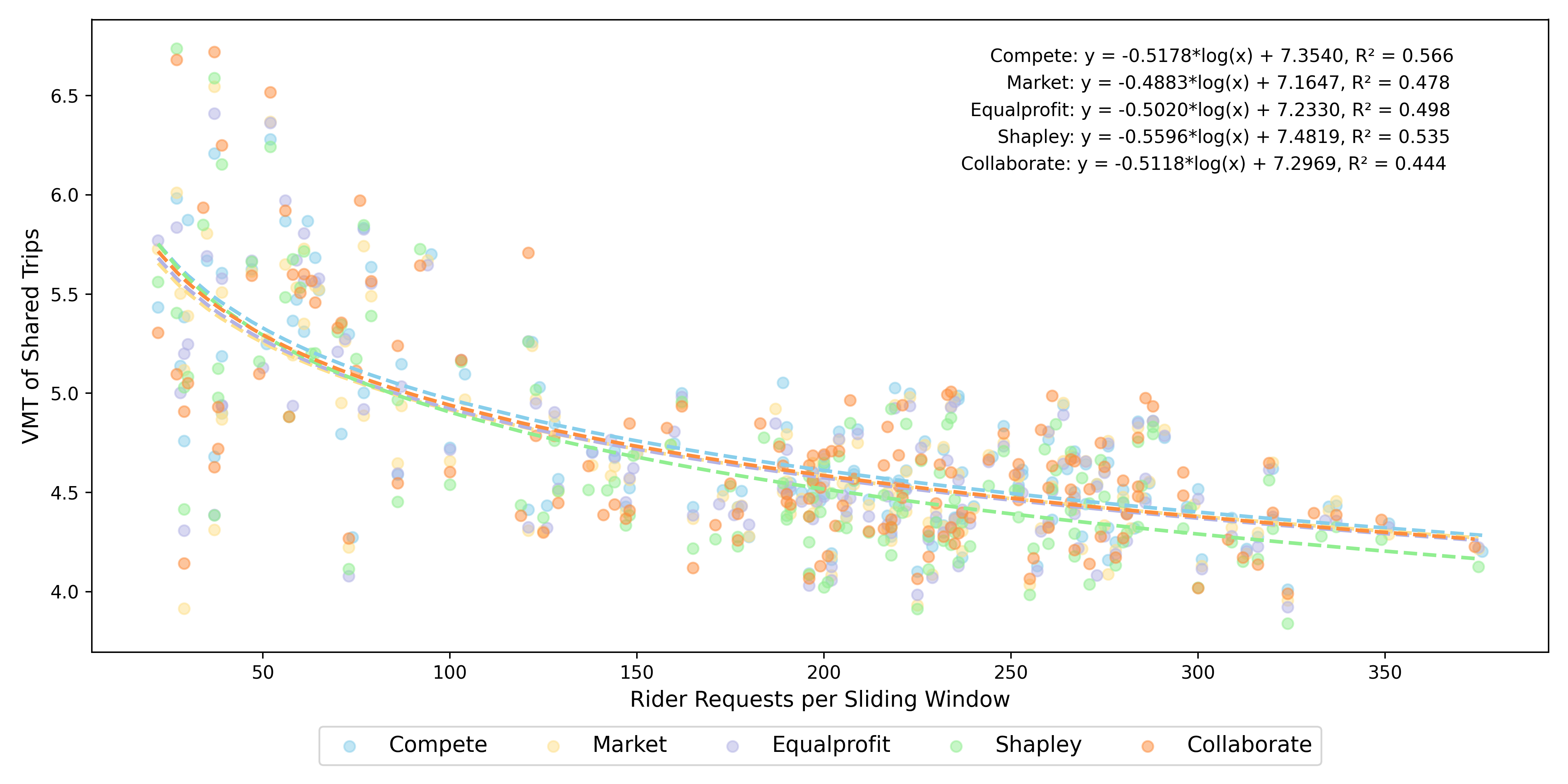}
    \caption{Performance Scaling: Vehicle Miles Traveled vs. Rider Requests per Sliding Window. All scenarios—competition, profit-aware collaboration (market share-based, equal profit-based, and Shapley-based mechanisms), and full collaboration—follow a sub-linear scaling with a consistent relationship.}
    \label{fig: vmt_scaling}
\end{figure}

The VMT of shared trips demonstrates a clear scaling behavior, consistently decreasing as the number of rider requests increases (Figure~\ref{fig: vmt_scaling}). 
The potential reason for this scaling behavior is that increased rider request density improves the likelihood of successfully matching riders traveling along similar routes, resulting in more efficient route consolidation.
Among the mechanisms examined, Shapley-value-based collaboration achieves the lowest VMT across most demand conditions, likely due to its contribution-based allocation strategy, which prioritizes globally efficient matches by evaluating each rider's marginal impact. This promotes consolidated routing and minimizes inefficient detours, particularly under dense request conditions. 

To quantify this relationship, we fitted a log-linear regression model of the form $\text{VMT} = \beta \log(N) + \alpha$, where $N$ denotes the number of rider requests per 10-minute interval. All mechanisms exhibit substantial logarithmic scaling, with $R^2$ values ranging from 0.444 (Full Collaboration) to 0.566 (Competition), indicating a strong and consistent relationship between rider request volume and VMT efficiency. (All coefficients are statistically significant with 99.9\% confidence.)

\subsection{Platform Profit} 
\begin{figure}[H]
    \centering
    \includegraphics[width=1\linewidth]{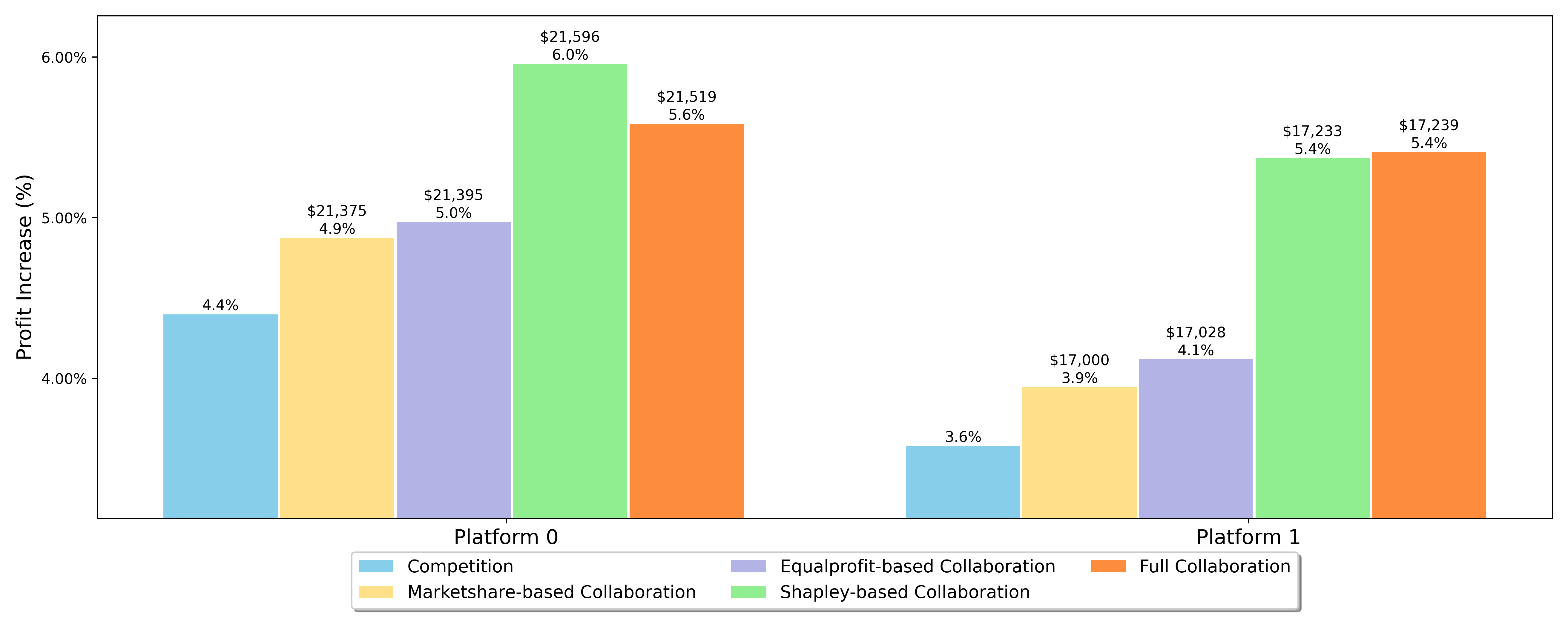}
    \caption{Profit Increase over Intra-platform and Cross-platform Ride-sharing}
    \label{fig: profit_increase}
\end{figure}

Figure~\ref{fig: profit_increase} presents the percentage improvement in profit each platform generates under various ride-sharing scenarios, relative to a baseline with neither ride-sharing nor collaboration. 
Firstly, ride-sharing proves beneficial even without collaboration between platforms. In the competitive setting, it raises the profits of Platform 0 and Platform 1 by 4.4\% and 3.6\%, respectively.
Cross-platform ride-sharing further improves profits over non-collaborative settings. The Shapley-value-based collaboration increases profits by 6.0\% for Platform~0 and 5.4\% for Platform~1, closely matching the outcomes under full collaboration (5.6\% and 5.4\%, respectively), and outperforming both the market-share-based and equal-profit-based approaches. 
This is because market-share and equal-profit schemes offer a static notion of fairness but overlook the marginal contribution of each platform in individual ride-sharing matches. When rewards diverge from actual contributions, platforms may perceive the outcome as unfair and be disincentivized from continuing collaboration. In contrast, fairness-aware mechanisms like the Shapley-value scheme allocate gains based on local, instance-level contributions. This alignment strengthens both incentive compatibility and voluntary participation, making these mechanisms more effective in fostering sustainable inter-platform cooperation.

\subsection{Riders’ Savings}
\begin{figure}[H]
    \centering
    \includegraphics[width=1\linewidth]{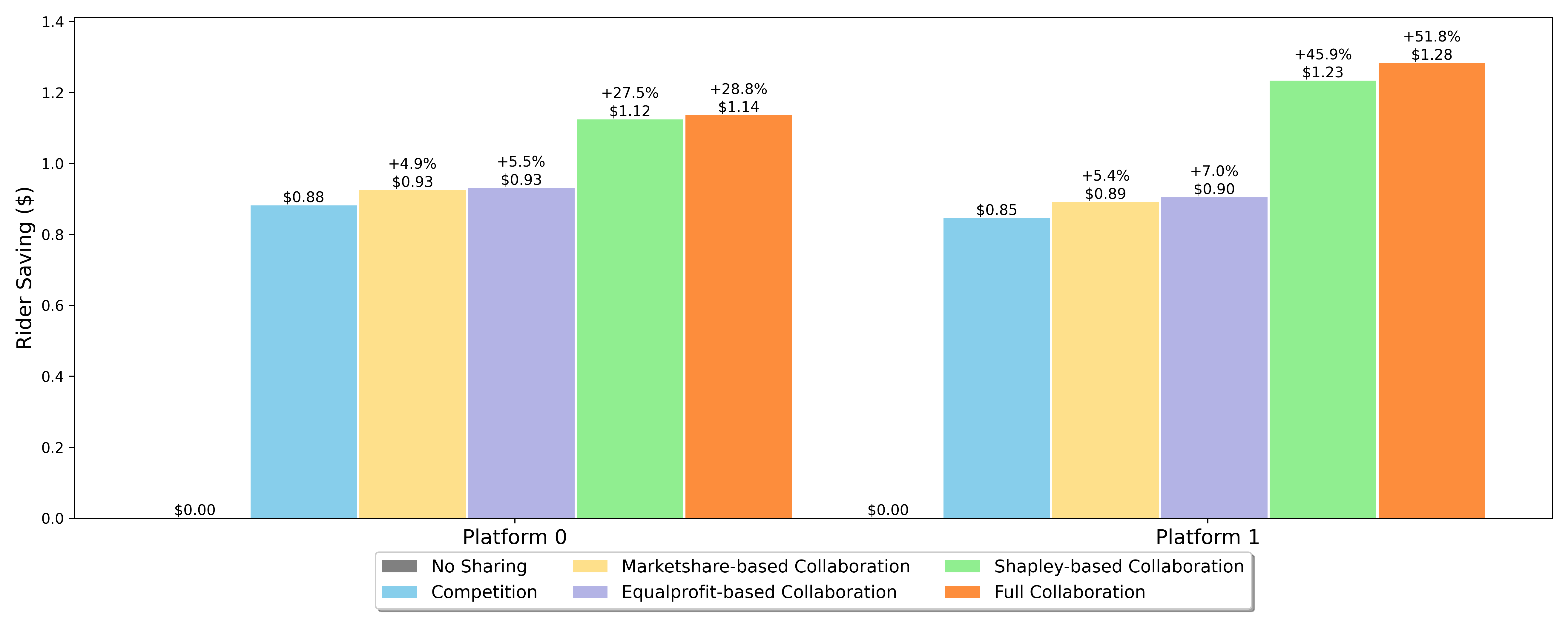}
    \caption{Average Rider Saving over Intra-platform and Cross-platform Ride-sharing}
    \label{fig: rider_saving}
\end{figure}

In our setting, riders receive a 10\% fare discount when successfully sharing a trip with another passenger. The resulting average savings per trip, relative to a zero-discount baseline, are shown in Figure~\ref{fig: rider_saving} across different scenarios.
While rider savings and platform profitability often represent competing objectives, the observed results suggest that collaboration may support improvements in both rider value and platform performance under certain conditions. 
However, this alignment may vary under different economic parameters, such as rider discount levels or driver commission rates, which can shift the balance of incentives across participants.
Even without inter-platform collaboration, ride-sharing yields meaningful cost reductions, with riders saving an average of \$0.88 on Platform~0 and \$0.85 per trip on Platform~1 under the competition scenario. Cross-platform collaboration further amplifies these gains. Under full collaboration, rider savings rise to \$1.14 and \$1.28 per trip for Platform~0 and Platform~1, corresponding to improvements of 28.8\% and 51.8\%, respectively, relative to the competitive ride-sharing baseline. The similarity in rider savings between the Shapley-value and market-share-based mechanisms may stem from their shared limitations in profit allocation, which fail to fully align with both platforms’ incentives or expectations.

\subsection{Passenger Detour Distance and Wait Time} 
\begin{figure}[htbp]
    \centering
    \includegraphics[width=1\linewidth]{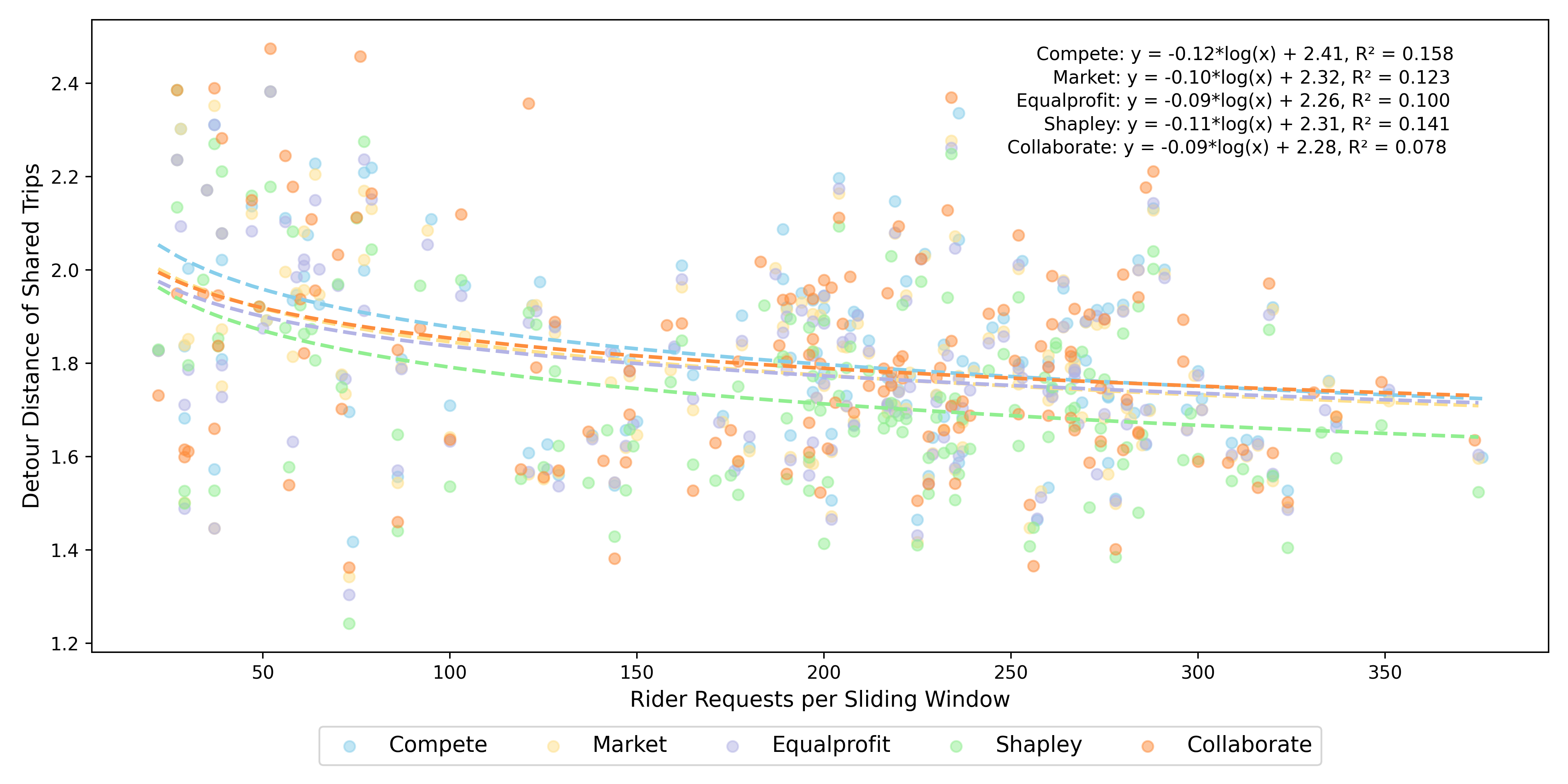}
    \caption{Performance Scaling: Detour Distance vs. Rider Requests per Sliding Window. All scenarios—competition, profit-aware collaboration (market share-based, equal profit-based, and Shapley-based mechanisms).}
    \label{fig: detour_scaling}
\end{figure}

\begin{figure}[H]
    \centering
    \includegraphics[width=1\linewidth]{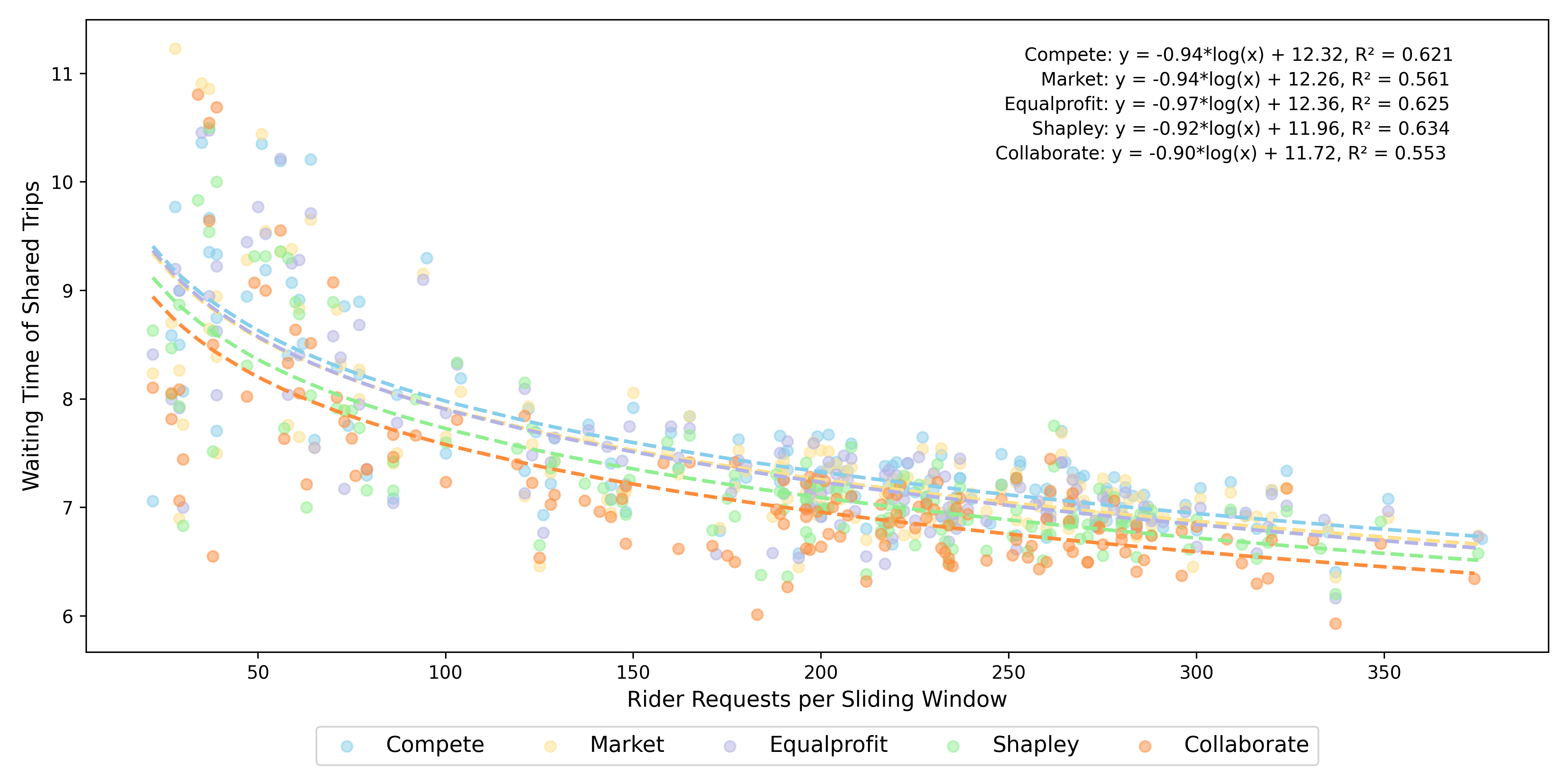}
    \caption{Performance Scaling: Wait Time vs. Rider Requests per Sliding Window. All scenarios—competition, profit-aware collaboration (market share-based, equal profit-based, and Shapley-based mechanisms).}
    \label{fig: waittime_scaling}
\end{figure}

Table~\ref{tab: wait_detour} reports the average passenger detour distance and wait time under different mechanisms, evaluated at a fixed request volume. Across all scenarios, collaboration improves both metrics compared to competition. Average detour distance ranges from 1.8106 miles (Competition) to 1.7268 miles (Shapley-value-based Collaboration), while wait times drop from 7.48 minutes (Competition) to 7.11 minutes (Full Collaboration). These results indicate that collaboration primarily contributes to reducing wait times, with relatively modest effects on detour distances. 
Figures~\ref{fig: detour_scaling} and~\ref{fig: waittime_scaling} extend the analysis by illustrating scaling performance under varying rider volumes. Each plot fits a log-linear model of the form $y = a\cdot\log(x) + b$, with shaded 95\% confidence intervals and reported $R^2$ values. A clear scaling trend is observed for wait time: all scenarios yield strong negative correlations with request volume, with $R^2$ values between 0.553 and 0.634. This suggests that higher demand improves the real-time matching rate and reduces dispatch latency, as a larger set of concurrent requests increases the probability of forming timely and spatially feasible matches. In contrast, detour distance shows limited sensitivity to request volume, with weaker fits ($R^2 < 0.16$ across all mechanisms). The weak correlation between demand and detour distance can be attributed to structural and behavioral constraints, which restrict the extent to which routes can be shared without exceeding passenger detour tolerance or operational feasibility limits.

\begin{table}[htbp]
\centering
\caption{Comparison of Service Quality Across Different Scenarios}
\label{tab: wait_detour} 
\begin{tabular}{@{}lcccc@{}}
\toprule
Scenario & Detour Distance (mile) & Wait Time (minute) \\
\midrule
Competition & 1.8106 & 7.48\\
Market-share-based Collaboration & 1.7847 & 7.41\\
Equal-profit-based Collaboration & 1.7811 & 7.39\\
Shapley-value-based Collaboration & 1.7268 & 7.24\\
Full Collaboration & 1.7980 & 7.11\\
\bottomrule
\end{tabular} 
\end{table} 

\section{Sensitivity Analysis} 
\label{sec: sensitivity}
This section evaluates whether the benefits of collaboration---particularly under the Shapley-based mechanism---remain robust across varying operational conditions. For clarity, we compare only three representative modes: competition, full collaboration, and Shapley-based profit-aware collaboration, as the latter was shown in the previous section to outperform the other profit-aware strategies with only marginal differences among them.

\subsection{Sensitivity Analysis Setup}
To evaluate the robustness of collaboration benefits, we conduct a sensitivity analysis by varying three key parameters that directly impact user behavior and platform economics: the fare discount factor ($\beta$), the maximum detour distance ($\gamma$), and the maximum rider waiting time ($\tau$). These are tested using a One-At-A-Time (OAT) approach while keeping other variables fixed (Table~\ref{tab:sensitivity_params}). 
In each experimental setting, we maintain a constant number of ride requests and enforce a fixed rider selection proportion of 58\% and 42\% for the two platforms. While the selections over platforms are regenerated in each replication to introduce stochasticity, the overall selection ratio is preserved. For each combination of parameters, five simulation replications are conducted to ensure the robustness and reliability of the results. A total of 50 simulation runs are conducted, covering all tested parameter values as summarized in Table~\ref{tab:sensitivity_params}.

\begin{table}[h]
\centering
\caption{Parameters varied in the sensitivity analysis}
\label{tab:sensitivity_params}
\begin{tabular}{lccc}
\hline
\textbf{Parameter} & \textbf{Symbol} & \textbf{Values Tested} & \textbf{Default Value} \\
\hline
Fare Discount Factor & $\beta$ & \{0.0, 0.1, 0.2, 0.3\} & 0.1 \\
Maximum Detour Factor & $\gamma$ & \{0.2, 0.4, 0.6, 0.8\} & 0.2 \\
Maximum Waiting Time (min) & $\tau$ & \{5, 10, 15, 20\} & 10 \\
\hline
\end{tabular}
\end{table}

\subsection{Sensitivity Analysis Results}
\begin{figure}[htbp]
    \centering
    \includegraphics[width=1\linewidth]{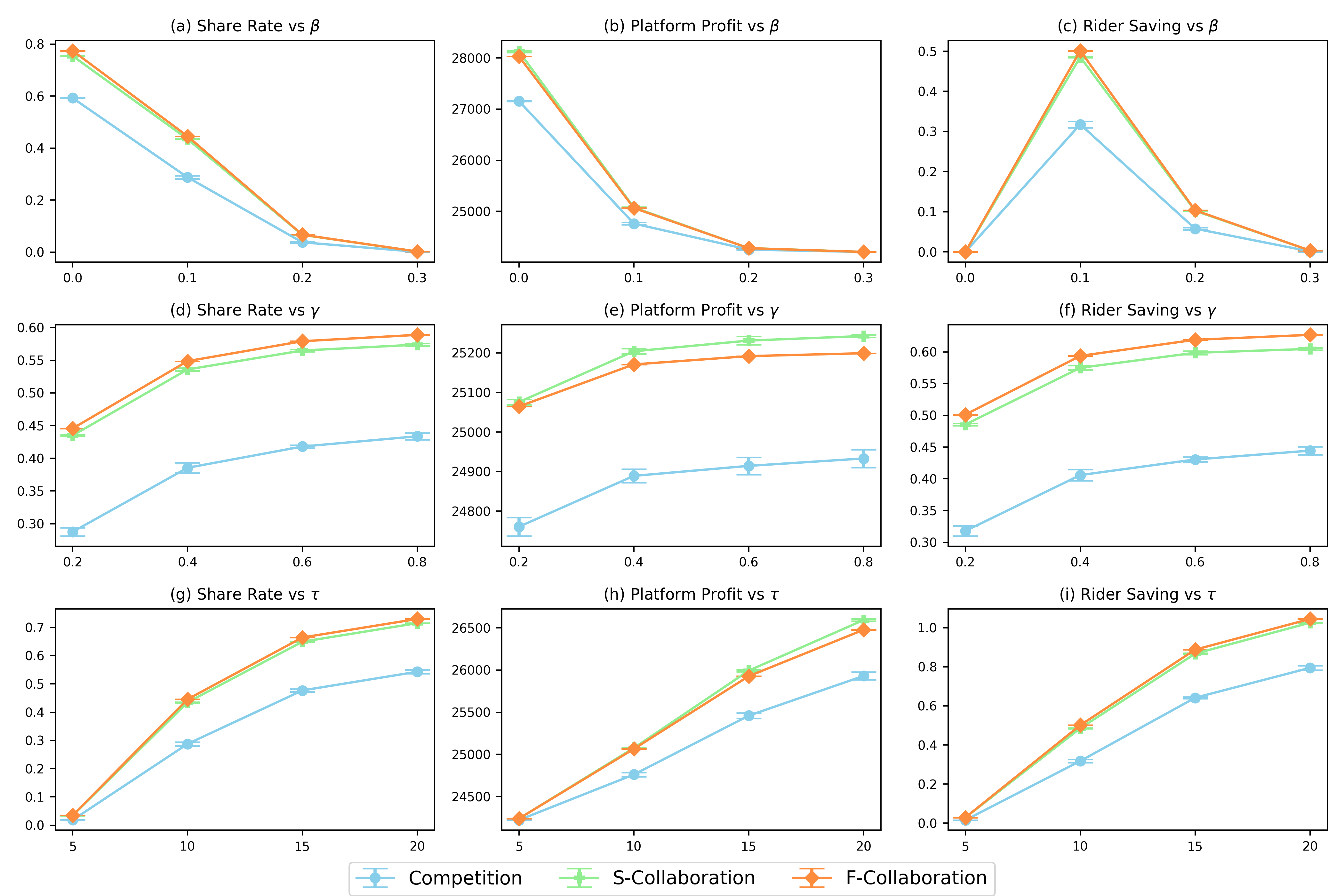}
    \caption{Sensitivity Analysis Results for Share Rate, Platform Profit, and Rider Savings over Discount Factor ($\beta$), Maximum Detour Factor($\gamma$), and Maximum Waiting Time ($\tau$)}
    \label{fig: Sensitivity_rate_profit}
\end{figure}

\begin{figure}[H]
    \centering
    \includegraphics[width=1\linewidth]{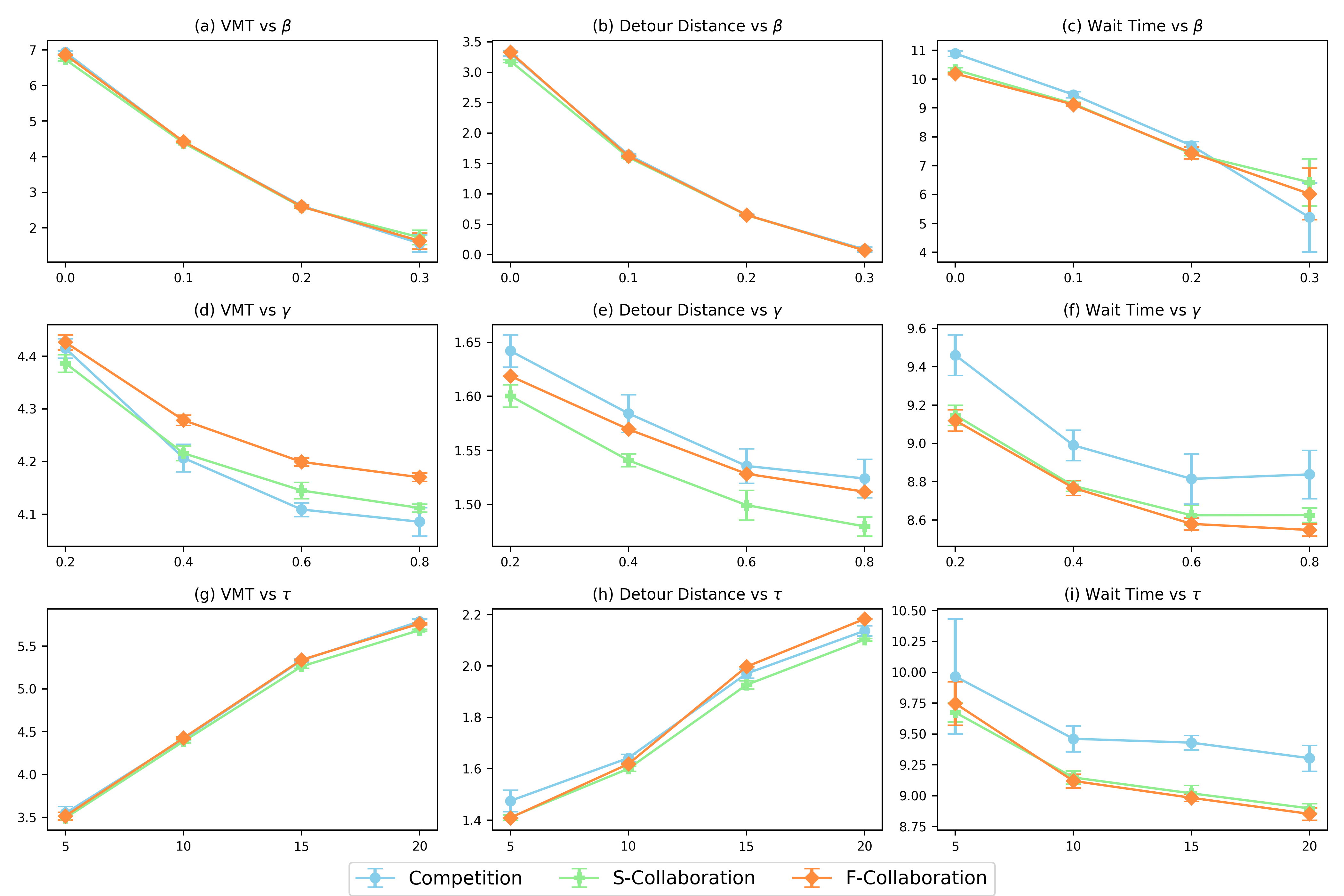}
    \caption{Sensitivity Analysis Results for VMT, Detour Distance and Waiting Time over Discount Factor ($\beta$), Maximum Detour Factor($\gamma$), and Maximum Waiting Time ($\tau$)}
    \label{fig: Sensitivity_distance_time}
\end{figure}

Figures~\ref{fig: Sensitivity_rate_profit} and~\ref{fig: Sensitivity_distance_time} present the sensitivity of key system performance metrics---share rate, platform profit, rider savings, vehicle miles traveled (VMT), passengers' detour distance, and waiting time---to variations in three parameters: fare discount factor ($\beta$), maximum detour factor ($\gamma$), and maximum waiting time ($\tau$). The results are shown under three scenarios labeled in the graphs: competition (Competition, blue), full collaboration (F-Collaboration, orange), and Shapley-value-based collaboration (S-Collaboration, green). Error bars indicate standard errors computed from five simulation replications per configuration.

Overall, the relative performance ranking among competition, full collaboration, and Shapley-value-based collaboration remains consistent across all tested parameter values, indicating the robustness of our baseline conclusions. In particular, Shapley-value-based collaboration consistently achieves high share rates, platform profits, and rider savings, close to or even better than the full collaboration scenario. This is because the Shapley-value-based mechanism allocates profits more fairly and proportionally based on each platform’s contribution to shared trips, which strengthens platform incentives to participate in cross-platform ride-sharing.

Providing a higher fare discount factor (i.e., higher $\beta$) leads to monotonic reductions in share rate, platform profit, vehicle miles traveled (VMT), detour distance and waiting time for the rider in shared trips, as shown in Figure~\ref{fig: Sensitivity_rate_profit}(a)–(b) and Figure~\ref{fig: Sensitivity_distance_time}(a)–(c). This is intuitive, as higher discounts reduce the net fare received by the platform, limiting profit margins and discouraging the acceptance of shared trips. The remaining budget can only support limited additional vehicle miles, detour distances, and, consequently, rider waiting times. In contrast, the average rider savings---calculated over all requests that opt in for ride-sharing, regardless of whether they are successfully matched---exhibit a non-monotonic trend: savings initially increase but decline as discounts become more generous, as presented in Figure~\ref{fig: Sensitivity_rate_profit}(c). As platform profit incentives diminish, fewer shared trips are accepted, and the resulting drop in share rate under high-discount-factor conditions reduces the average rider savings, even though individual matched riders still benefit.

Authorizing greater detour upper bounds (i.e., higher $\gamma$) increases the share rate and enhances ride-sharing efficiency (reductions in VMT, detour distance, and wait time), as evidenced by Figure~\ref{fig: Sensitivity_rate_profit}(d)–(f) and Figure~\ref{fig: Sensitivity_distance_time}(d)–(f). This is because looser detour constraints expand the feasible matching space, enabling the system to pair more riders whose routes partially overlap. However, the marginal benefits diminish beyond $\gamma = 0.6$, as the newly added ride-sharing pairs inherently involve longer detours.
In such cases, the additional flexibility primarily permits low-quality matches that offer limited contributions to overall system performance.

Similarly, raising the waiting time upper bound ($\tau$) improves share rates and economic outcomes such as platform profit and rider savings, as seen in Figure~\ref{fig: Sensitivity_rate_profit}(g)–(i). This is because a higher tolerance for waiting enables the system to consider a broader set of rider pairs within a longer temporal window, thereby increasing the likelihood of successful matches. However, this additional flexibility also leads to longer detours and higher vehicle miles traveled (VMT), as shown in Figure~\ref{fig: Sensitivity_distance_time}(g)–(h). A possible explanation is that the expanded matching window admits more rider pairs that are profitable but spatially dispersed. Since the system employs a maximum weighted matching algorithm based on platform profit, it may favor combinations with high economic returns even if they result in longer shared routes, thereby increasing VMT under our definition. Interestingly, although the algorithm permits longer waiting times, the actual wait times of shared trips do not increase significantly. This is because effective matching decisions are jointly constrained by other parameters, such as the maximum detour factor, which limits how much temporal flexibility can be utilized in practice.

\newpage
\section{Conclusion}
\label{sec: conclusion} 
This study investigated the benefits of cross-platform ride-sharing under various profit allocation mechanisms. By modeling rider–rider graphs and applying a profit-aware maximum-weighted matching algorithm, we evaluated the performance of three profit-sharing schemes: equal-profit, market-share-based, and Shapley-value-based. Our findings indicate that the Shapley-value-based collaboration consistently yields superior system-level outcomes, including higher share rates, increased platform profits, enhanced rider savings, and reduced waiting times. These advantages persist across diverse operational settings, demonstrating the mechanism's scalability and adaptability. Furthermore, we observed clear evidence of economies of scale: as the number of rider requests increases, share rates improve, and total vehicle miles traveled (VMT) per shared trip decreases. While detour distances remain relatively stable with weak correlation to request volume, average wait times tend to decline modestly with sufficient vehicle supply. These economies of scale suggest that larger-scale deployments can enhance overall system efficiency and passenger experience under favorable operating conditions.

Complementing these findings, we identified structural scaling patterns—super-linear edge growth and sub-linear average degree scaling—that elucidate the system’s increasing efficiency with scale (see Appendix). Super-linear edge growth indicates that inter-platform collaboration substantially increases network density even at small scales. In contrast, sub-linear degree scaling implies that each rider experiences only a modest increase in potential matches as the system grows. Additionally, regression models derived from our simulation results not only capture performance trends but also serve as practical tools for forecasting system scalability. These insights can inform adaptive deployment strategies in distributed systems, where computational tasks such as graph construction and matching are parallelizable.

It is important to note that this study primarily focuses on the rider side and assumes sufficient vehicle supply, without modeling under-supply scenarios that could impact performance in various spatial or temporal contexts. Moreover, the current framework supports only two-person ride-sharing; multi-rider pooling and its associated scheduling complexities are not considered. The simulation operates under a centralized setting, without addressing the computational scalability required for large-scale, distributed deployments. Furthermore, rider platform choice is treated as exogenous, without modeling behavioral responses to fare discounts or service attributes.

Future research could address these limitations in several directions. First, investigating operational strategies under constrained vehicle supply—either system-wide, at specific times, or in localized areas—would assess platform robustness and efficiency. Second, extending the framework to support multi-rider pooling would enhance its applicability, particularly in high-density settings where three or more riders can feasibly share a trip. This extension would necessitate revisiting graph construction and allocation rules to accommodate higher-order groupings. Third, exploring scalable computational strategies, such as trip-based graph partitioning, could enable parallelized matching on large-scale graphs. Finally, incorporating an endogenous mode choice model—explicitly capturing how riders respond to fare discounts and platform characteristics—could provide deeper insights into the trade-offs between incentive design and network-level outcomes.

\section{Acknowledgements}
This research was supported by the NSF Grant CMMI-2052337.

\section{Author Contributions}
The authors confirm their contribution to the paper as follows: study conception and design: X. Dong, J. Ventura, V. Gayah; simulation: X. Dong; analysis and interpretation of results: X. Dong, J. Ventura, V. Gayah; draft manuscript preparation: X. Dong, J. Ventura, V. Gayah; All authors reviewed the results and approved the final version of the manuscript.

\newpage
\section*{Appendices: Densification Patterns Observed on the Growth of Rider-Rider Graphs}
\label{sec: scale law}
Beyond evaluating system performance, we empirically examine the structural evolution of the rider-rider graph as the system scales. Our results reveal two consistent scaling patterns: super-linear edge growth and sub-linear scaling of the average graph degree. These structural properties suggest that as the number of ride requests increases, the matching graph becomes denser without a proportional increase in local connectivity, helping to explain the scalability and efficiency of our framework.

\subsection*{A. Super-linear Scaling of Edge Growth}
\begin{figure}[H]
    \centering
    \includegraphics[width=0.8\linewidth]{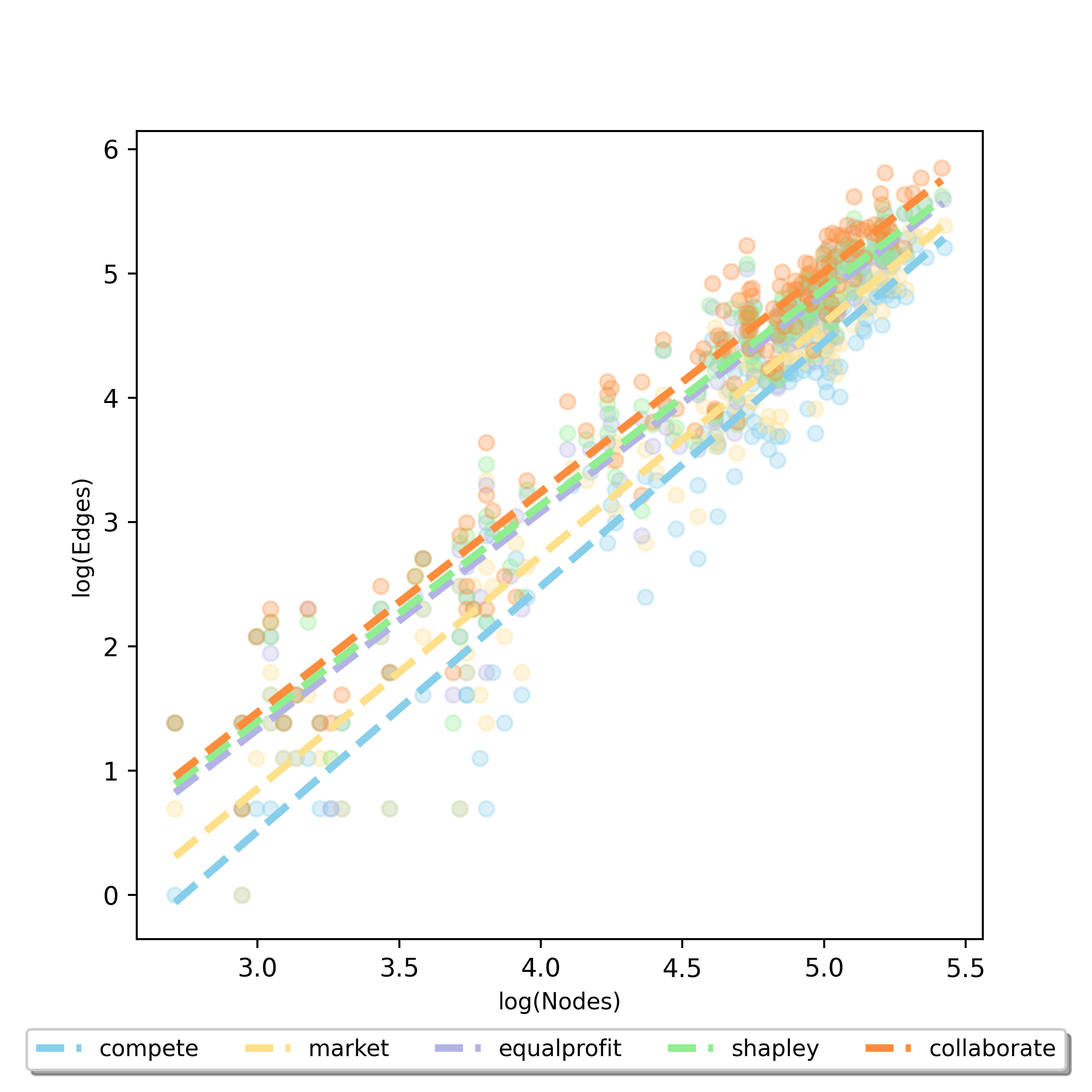}
    \caption{Edge Growth vs. Nodes in log-log scales. All scenarios—competition, profit-aware collaboration (market share-based, equal profit-based, and Shapley-based mechanisms), and full collaboration—follow the Densification Power Law with a consistently strong fit.}
    \label{fig: DPL_networkscaling}
\end{figure}

Figure~\ref{fig: DPL_networkscaling} presents the log-log relationship between the number of nodes (trip requests) and the number of edges (feasible ride-sharing pairs) across various scenarios. First of all, all scenarios exhibit strong linear trends, confirming a power-law relationship between network size and connectivity. The estimated scaling exponents $\beta$ range from 1.7371 to 1.9661 (Table~\ref{tab:log_edges_vs_log_nodes}), indicating that edge growth outpaces node growth as the system scales. 
This observation is consistent with the Densification Power Law identified by Leskovec et al.~\citep{leskovec2005graphs}, which states that in many real-world networks, the number of edges grows super-linearly with the number of nodes, following the relationship $|E| \propto |V|^{\alpha}$ with $1 < \alpha < 2$.
In the context of ride-sharing, such densification implies that as the system grows, more potential matches become available, which in turn increases the likelihood of successful pairings, enhances scheduling flexibility, and improves overall operational efficiency.

Beyond this general growth pattern, the vertical positioning of the regression lines shows that collaboration-based mechanisms consistently produce more feasible matches than the competition model at the same request volume. This advantage arises from inter-platform matching, which expands the feasible pairing space.
While the competition scenario exhibits the steepest exponent ($\beta = 1.9661$), it consistently yields fewer edges than collaborative mechanisms across the observed range. 
From a mechanism design perspective, collaboration models are constructed by adding inter-platform links to the competition model’s intra-platform graph, making their feasible match sets strict supersets. As such, the competition scenario is structurally unlikely to ever outperform collaboration. While regression suggests a potential crossover at $N \approx 955$, where competition could exceed Shapley in edge count, this lies far beyond the observed range (up to 200 shared trip requests per 5 minutes), and model refitting would be required to assess behavior at such scales. 

\begin{table}[htbp]
    \centering
    \caption{Regression Results: $\log(\text{Edges}) \sim \log(\text{Nodes})$}
    \label{tab:log_edges_vs_log_nodes}
    \begin{tabular}[width = \linewidth]{l
                    S[table-format=1.4]
                    S[table-format=-1.4]
                    c
                    c}
        \toprule
        \textbf{Mechanism} & {$\beta$ (Coefficient)} & {$\alpha$ (Intercept)} & {$R^2$} & P-value \\
        \midrule
        Competition        & 1.9661  & -5.3839 & 0.9109 & $p < 0.001$*** \\
        Market-share-based Collaboration         & 1.8726  & -4.7619 & 0.9185 & $p < 0.001$*** \\
        Equal-profit-based Collaboration    & 1.7488  & -3.9132 & 0.9281 & $p < 0.001$*** \\
        Shapley-value-based Collaboration        & 1.7371  & -3.8128 & 0.9306 & $p < 0.001$*** \\
        Full Collaborate    & 1.7719  & -3.8464 & 0.9285 & $p < 0.001$*** \\
        \bottomrule
    \end{tabular}
    
    \vspace{0.3cm}
    \footnotesize
    \textit{Note:} The model is specified as $\log(\text{Edges}) = \beta \log(\text{Nodes}) + \alpha$. *** is significant at the 0.1\% significance level.
\end{table}

\subsection*{B. Sub-linear Scaling of Average Graph Degree}
\begin{figure}[H]
    \centering
    \includegraphics[width=0.8\linewidth]{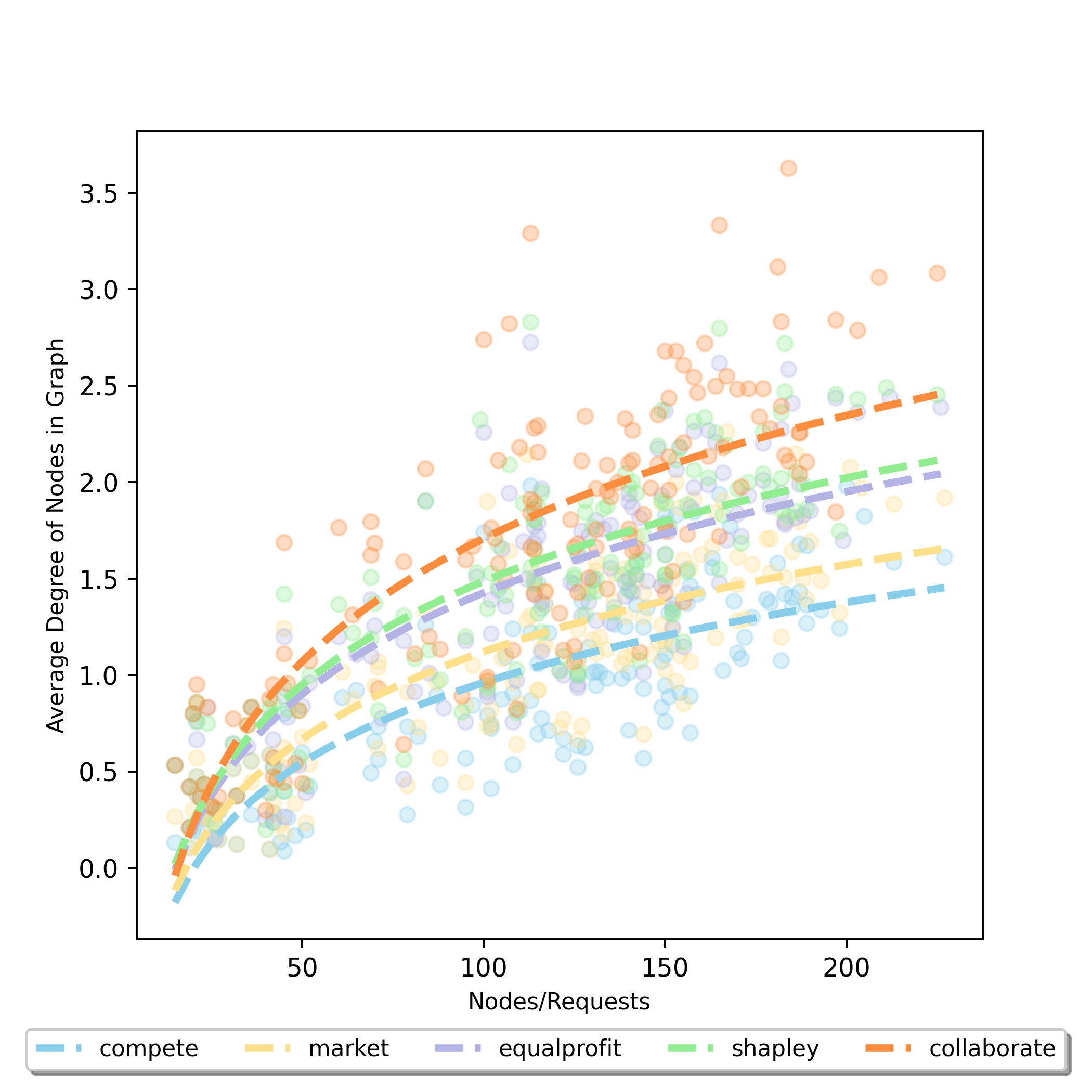}
    \caption{Average Degree of Graph vs. Nodes. All scenarios—competition, profit-aware collaboration (market share-based, equal profit-based, and Shapley-based mechanisms), and full collaboration—follow a sub-linear scaling pattern with a consistently strong fit.}
    \label{fig: degree_nodes}
\end{figure}

While the previous section demonstrates that the number of edges grows super-linearly with the number of nodes, this does not imply a similar trend in average degree. Instead of capturing overall network expansion, the average degree reflects local connectivity, measuring the expected number of feasible matches per rider. For example, an average degree of 2 indicates that each rider has, on average, two potential ride-sharing partners.
We model the relationship between average degree $d$ and the number of riders $N$ using a log-linear form: $d = \beta \log(N) + \alpha$. As shown in Table~\ref{tab:degrees_vs_nodes} and Figure~\ref{fig: degree_nodes}, all scenarios exhibit clear logarithmic growth with strong fit results ($R^2 > 0.6$), indicating that rider-level connectivity improves with system size, but at a sub-linear rate. This trend is consistent with observations in network evolution~\citep{leskovec2005graphs}, where the average degree increases more slowly than edge count as networks densify.

A meaningful threshold occurs when the average degree reaches 1, indicating that each rider, on average, has at least one potential match. Below this level, the network remains too sparse to support consistent sharing. Under collaborative mechanisms, this threshold is typically reached with around 50 riders per matching window, while the competition model requires over 100 riders per matching window. This emphasizes the advantage of inter-platform collaboration in enabling viable ride-sharing at smaller operational scales.
Among collaborative mechanisms, full collaboration---which permits unrestricted inter-platform matching---achieves the highest average degrees across all system sizes. In contrast, profit-aware strategies such as equal profit and Shapley result in slightly lower average degrees. This is because their allocation rules may be perceived as unfavorable by one of the platforms involved, leading them to reject otherwise feasible inter-platform matches and degrade overall system performance. These findings reinforce that well-designed profit-sharing mechanisms are essential to unlocking the full potential of platform collaboration.


\begin{table}[H]
    \centering
    \caption{Regression Results: $\text{graph average degree} \sim \log(\text{Nodes})$}
    \label{tab:degrees_vs_nodes}
    \begin{tabular}{l
                    S[table-format=1.6]
                    S[table-format=-1.6]
                    S[table-format=1.6]
                    c}
        \toprule
        \textbf{Mechanism} & {$\beta$ (Coefficient)} & {$\alpha$ (Intercept)} & {$R^2$} & P-value \\
        \midrule
        Competition        & 0.600697  & -1.805302 & 0.613997 & $p < 0.001$*** \\
        Market-share-based Collaboration         & 0.652485  & -1.883932 & 0.643614 & $p < 0.001$*** \\
        Equal-profit-based Collaboration   & 0.757854  & -2.064280 & 0.652920 & $p < 0.001$*** \\
        Shapley-value-based Collaboration        & 0.774063  & -2.078778 & 0.654131 & $p < 0.001$*** \\
        Full Collaboration    & 0.921266  & -2.534843 & 0.625059 & $p < 0.001$*** \\
        \bottomrule
    \end{tabular}
    
    \vspace{0.3cm}
    \footnotesize
    \textit{Note:} The model is specified as $\text{graph average degree} = \beta log(x) + \alpha$. *** is significant at the 0.1\% significance level.
\end{table}

\subsection*{C. Computational Efficiency}
\label{sec: computation time}
\begin{figure}[htbp]
    \centering
    \includegraphics[width=1\linewidth]{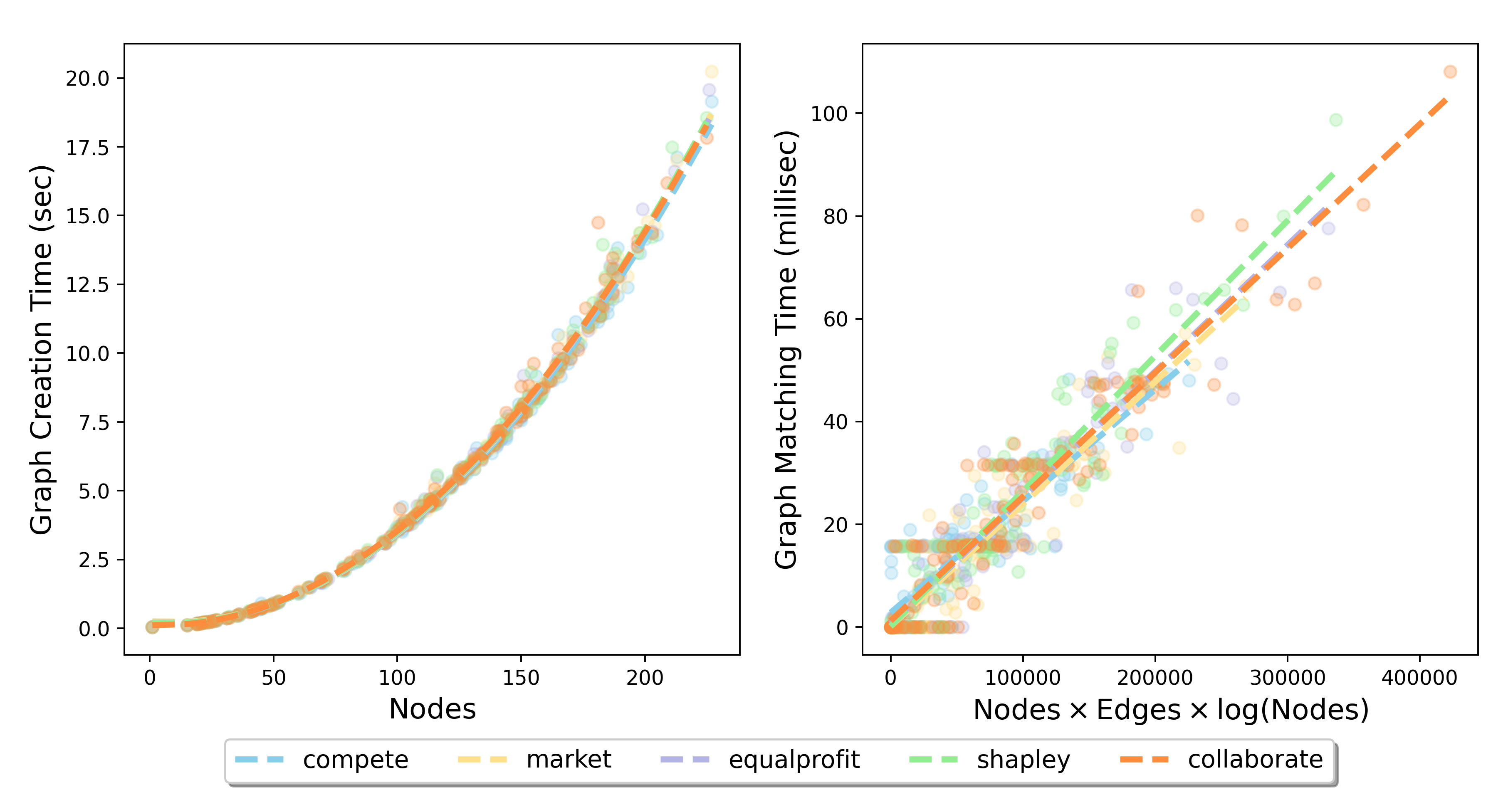}
    \caption{Graph Creation Time and Matching Time}
    \label{fig: computation time}
\end{figure}

While ILP provides a general modeling framework for ride-sharing optimization, the underlying problem is NP-Hard, and its reliance on generic solvers and the lack of exploitable problem structure often lead to scalability bottlenecks in large-scale settings~\cite{alonso2017demand}. In contrast, our graph-based framework leverages inherent structural properties to enable both theoretical insights and practical efficiency, achieving polynomial-time complexity in both graph construction and matching.
Our result not only reveals key scaling laws—such as edge growth, node degree behavior, and system-level performance—but also facilitates tractable and efficient computation via streamlined graph construction and faster matching algorithms.

Figure~\ref{fig: computation time} illustrates the computational costs of graph construction and matching under various coordination mechanisms. Graph creation time scales quadratically with the number of nodes ($t_c \sim |V|^2$), corresponding to the rider population. Regression results across all mechanisms show excellent fit ($R^2 > 0.99$; Table~\ref{tab:creation_time_vs_nodes}), confirming the predictability and consistency of construction time.
Graph matching time is significantly lower and exhibits a near-linear relationship with a composite metric involving node and edge counts, i.e., $t_m \sim |V||E|\log(|V|)$—consistent with the theoretical complexity of efficient matching algorithms such as those described in~\cite{galil1986efficient}. Regression outcomes in Table~\ref{tab:matching_time_vs_nodes} further validate this pattern with high explanatory power ($R^2 > 0.89$ across mechanisms).
Among all mechanisms, the Shapley-based scheme produces the most connected graphs—approaching full collaboration—thus enabling broader cross-platform ride-sharing. However, it incurs the highest matching time, reflecting a trade-off between fairness and computational cost. In contrast, the market share-based mechanism leads to sparser graphs with fewer inter-platform edges and lower runtime. 
Importantly, the total computation time (including the graph creation and matching time) remains consistently under 5 seconds in our tested system sizes, demonstrating the viability of deploying our approach in real-time settings.

\begin{table}[htbp]
    \centering
    \caption{Regression Results: Graph Creation Time vs Number of Nodes}
    \label{tab:creation_time_vs_nodes}
    \begin{tabular}{l
                    S[table-format=1.4]
                    S[table-format=-1.4]
                    S[table-format=1.6]
                    cc}
        \toprule
        \textbf{Mechanism} & {$\beta_2$ (Quadratic)} & {$\beta_1$ (Linear)} & {$\alpha$ (Intercept)} & {$R^2$} & P-value \\
        \midrule
        Competition        & 0.000374  & -0.004947 & 0.184042 & 0.995019 & $p < 0.001$*** \\
        Market-share-based Collaboration         & 0.000388  & -0.006885 & 0.258802 & 0.996828 & $p < 0.001$*** \\
        Equal-profit-based Collaboration   & 0.000383  & -0.005590 & 0.208667 & 0.996699 & $p < 0.001$*** \\
        Shapley-value-based Collaboration        & 0.000384  & -0.005371 & 0.199609 & 0.994679 & $p < 0.001$*** \\
        Full Collaboration    & 0.000372  & -0.002897 & 0.116054 & 0.992847 & $p < 0.001$*** \\
        \bottomrule
    \end{tabular}
    
    \vspace{0.3cm}
    \footnotesize
    \textit{Note:} The model is specified as $\text{Creation Time} = \beta_2 x^2 + \beta_1 x + \alpha$. *** is significant at the 0.1\% significance level.
\end{table}

\begin{table}[htbp]
    \centering
    \caption{Regression Results: Graph Matching Time vs Number of Nodes}
    \label{tab:matching_time_vs_nodes}
    \begin{tabular}{l
                    S[table-format=1.6]
                    S[table-format=-1.4]
                    S[table-format=1.6]
                    c}
        \toprule
        \textbf{Mechanism} & {$\beta$ (Coefficient)} & {$\alpha$ (Intercept)} & {$R^2$} & P-value \\
        \midrule
        Competition        & 0.000217  & 2.8106 & 0.7685 & $p < 0.001$*** \\
        Market-share-based Collaboration         & 0.000238  & 0.1583 & 0.8582 & $p < 0.001$*** \\
        Equal-profit-based Collaboration   & 0.000246  & 0.6334 & 0.8947 & $p < 0.001$*** \\
        Shapley-value-based Collaboration        & 0.000263  & 0.1380 & 0.9076 & $p < 0.001$*** \\
        Full Collaboration    & 0.000242  & 1.2146 & 0.9066 & $p < 0.001$*** \\
        \bottomrule
    \end{tabular}
    
    \vspace{0.3cm}
    \footnotesize
    \textit{Note:} The model is specified as $\text{Matching Time} = \beta x + \alpha$. *** is significant at the 0.1\% significance level.
\end{table}

\newpage
\bibliographystyle{elsarticle-harv}
\bibliography{main}

\begin{thebibliography}{40}
\expandafter\ifx\csname natexlab\endcsname\relax\def\natexlab#1{#1}\fi
\providecommand{\url}[1]{\texttt{#1}}
\providecommand{\href}[2]{#2}
\providecommand{\path}[1]{#1}
\providecommand{\DOIprefix}{doi:}
\providecommand{\ArXivprefix}{arXiv:}
\providecommand{\URLprefix}{URL: }
\providecommand{\Pubmedprefix}{pmid:}
\providecommand{\doi}[1]{\href{http://dx.doi.org/#1}{\path{#1}}}
\providecommand{\Pubmed}[1]{\href{pmid:#1}{\path{#1}}}
\providecommand{\bibinfo}[2]{#2}
\ifx\xfnm\relax \def\xfnm[#1]{\unskip,\space#1}\fi
\bibitem[{Alonso-Mora et~al.(2017)Alonso-Mora, Samaranayake, Wallar, Frazzoli and Rus}]{alonso2017demand}
\bibinfo{author}{Alonso-Mora, J.}, \bibinfo{author}{Samaranayake, S.}, \bibinfo{author}{Wallar, A.}, \bibinfo{author}{Frazzoli, E.}, \bibinfo{author}{Rus, D.}, \bibinfo{year}{2017}.
\newblock \bibinfo{title}{On-demand high-capacity ride-sharing via dynamic trip-vehicle assignment}.
\newblock \bibinfo{journal}{Proceedings of the National Academy of Sciences} \bibinfo{volume}{114}, \bibinfo{pages}{462--467}.
\bibitem[{Boeing(2017)}]{boeing2017osmnx}
\bibinfo{author}{Boeing, G.}, \bibinfo{year}{2017}.
\newblock \bibinfo{title}{Osmnx: New methods for acquiring, constructing, analyzing, and visualizing complex street networks}.
\newblock \bibinfo{journal}{Computers, environment and urban systems} \bibinfo{volume}{65}, \bibinfo{pages}{126--139}.
\bibitem[{Chen et~al.(2023)Chen, Ke and Chen}]{chen2023quantifying}
\bibinfo{author}{Chen, W.}, \bibinfo{author}{Ke, J.}, \bibinfo{author}{Chen, X.}, \bibinfo{year}{2023}.
\newblock \bibinfo{title}{Quantifying traffic emission reductions and traffic congestion alleviation from high-capacity ride-sharing}.
\newblock \bibinfo{journal}{arXiv preprint arXiv:2308.10512} .
\bibitem[{Chen et~al.(2025)Chen, Yang, Chen and Ke}]{chen2025scaling}
\bibinfo{author}{Chen, W.}, \bibinfo{author}{Yang, L.}, \bibinfo{author}{Chen, X.}, \bibinfo{author}{Ke, J.}, \bibinfo{year}{2025}.
\newblock \bibinfo{title}{Scaling laws of dynamic high-capacity ride-sharing}.
\newblock \bibinfo{journal}{Transportation Research Part C: Emerging Technologies} \bibinfo{volume}{174}, \bibinfo{pages}{105064}.
\bibitem[{Datar et~al.(2002)Datar, Gionis, Indyk and Motwani}]{datar2002maintaining}
\bibinfo{author}{Datar, M.}, \bibinfo{author}{Gionis, A.}, \bibinfo{author}{Indyk, P.}, \bibinfo{author}{Motwani, R.}, \bibinfo{year}{2002}.
\newblock \bibinfo{title}{Maintaining stream statistics over sliding windows}.
\newblock \bibinfo{journal}{SIAM Journal on Computing} \bibinfo{volume}{31}, \bibinfo{pages}{1794--1813}.
\bibitem[{Dijkstra(2022)}]{dijkstra2022note}
\bibinfo{author}{Dijkstra, E.W.}, \bibinfo{year}{2022}.
\newblock \bibinfo{title}{A note on two problems in connexion with graphs}, in: \bibinfo{booktitle}{Edsger Wybe Dijkstra: his life, work, and legacy}, pp. \bibinfo{pages}{287--290}.
\bibitem[{Dong et~al.(2024)Dong, Liu and Gayah}]{dong2024analytical}
\bibinfo{author}{Dong, X.}, \bibinfo{author}{Liu, H.}, \bibinfo{author}{Gayah, V.V.}, \bibinfo{year}{2024}.
\newblock \bibinfo{title}{An analytical model of many-to-one carpool system performance under cost-based detour limits}.
\newblock \bibinfo{journal}{International Journal of Transportation Science and Technology} .
\bibitem[{Edmonds(1965)}]{edmonds1965maximum}
\bibinfo{author}{Edmonds, J.}, \bibinfo{year}{1965}.
\newblock \bibinfo{title}{Maximum matching and a polyhedron with 0, 1-vertices}.
\newblock \bibinfo{journal}{Journal of research of the National Bureau of Standards B} \bibinfo{volume}{69}, \bibinfo{pages}{55--56}.
\bibitem[{Ferguson(1997)}]{ferguson1997rise}
\bibinfo{author}{Ferguson, E.}, \bibinfo{year}{1997}.
\newblock \bibinfo{title}{The rise and fall of the american carpool: 1970--1990}.
\newblock \bibinfo{journal}{Transportation} \bibinfo{volume}{24}, \bibinfo{pages}{349--376}.
\bibitem[{Fielbaum et~al.(2021)Fielbaum, Bai and Alonso-Mora}]{fielbaum2021demand}
\bibinfo{author}{Fielbaum, A.}, \bibinfo{author}{Bai, X.}, \bibinfo{author}{Alonso-Mora, J.}, \bibinfo{year}{2021}.
\newblock \bibinfo{title}{On-demand ridesharing with optimized pick-up and drop-off walking locations}.
\newblock \bibinfo{journal}{Transportation research part C: emerging technologies} \bibinfo{volume}{126}, \bibinfo{pages}{103061}.
\bibitem[{Galil(1986)}]{galil1986efficient}
\bibinfo{author}{Galil, Z.}, \bibinfo{year}{1986}.
\newblock \bibinfo{title}{Efficient algorithms for finding maximum matching in graphs}.
\newblock \bibinfo{journal}{ACM Computing Surveys (CSUR)} \bibinfo{volume}{18}, \bibinfo{pages}{23--38}.
\bibitem[{Guo et~al.(2023)Guo, Qu, Zhang, Noursalehi and Zhao}]{GUO2023104397}
\bibinfo{author}{Guo, X.}, \bibinfo{author}{Qu, A.}, \bibinfo{author}{Zhang, H.}, \bibinfo{author}{Noursalehi, P.}, \bibinfo{author}{Zhao, J.}, \bibinfo{year}{2023}.
\newblock \bibinfo{title}{Dissolving the segmentation of a shared mobility market: A framework and four market structure designs}.
\newblock \bibinfo{journal}{Transportation Research Part C: Emerging Technologies} \bibinfo{volume}{157}, \bibinfo{pages}{104397}.
\bibitem[{Heineke et~al.(2021)Heineke, Kloss, M{\"o}ller and Wiemuth}]{heineke2021shared}
\bibinfo{author}{Heineke, K.}, \bibinfo{author}{Kloss, B.}, \bibinfo{author}{M{\"o}ller, T.}, \bibinfo{author}{Wiemuth, C.}, \bibinfo{year}{2021}.
\newblock \bibinfo{title}{Shared mobility: Where it stands and where it’s going}.
\newblock \bibinfo{journal}{McKinsey Center for Future Mobility} .
\bibitem[{Jordahl(2014)}]{jordahl2014geopandas}
\bibinfo{author}{Jordahl, K.}, \bibinfo{year}{2014}.
\newblock \bibinfo{title}{Geopandas: Python tools for geographic data}.
\newblock \bibinfo{journal}{URL: https://github. com/geopandas/geopandas} .
\bibitem[{Ke et~al.(2024)Ke, Wang, Masoud, Schiffer and Correia}]{ke2024emerging}
\bibinfo{author}{Ke, J.}, \bibinfo{author}{Wang, H.}, \bibinfo{author}{Masoud, N.}, \bibinfo{author}{Schiffer, M.}, \bibinfo{author}{Correia, G.H.}, \bibinfo{year}{2024}.
\newblock \bibinfo{title}{Emerging on-demand passenger and logistics systems: Modelling, optimization, and data analytics}.
\bibitem[{Ke and Qian(2023)}]{ke2023leveraging}
\bibinfo{author}{Ke, Z.}, \bibinfo{author}{Qian, S.}, \bibinfo{year}{2023}.
\newblock \bibinfo{title}{Leveraging ride-hailing services for social good: Fleet optimal routing and system optimal pricing}.
\newblock \bibinfo{journal}{Transportation Research Part C: Emerging Technologies} \bibinfo{volume}{155}, \bibinfo{pages}{104284}.
\bibitem[{Lehe et~al.(2021)Lehe, Gayah and Pandey}]{lehe2021increasing}
\bibinfo{author}{Lehe, L.}, \bibinfo{author}{Gayah, V.V.}, \bibinfo{author}{Pandey, A.}, \bibinfo{year}{2021}.
\newblock \bibinfo{title}{Increasing returns to scale in carpool matching: Evidence from scoop}.
\newblock \bibinfo{journal}{Transport findings} .
\bibitem[{Leskovec et~al.(2005)Leskovec, Kleinberg and Faloutsos}]{leskovec2005graphs}
\bibinfo{author}{Leskovec, J.}, \bibinfo{author}{Kleinberg, J.}, \bibinfo{author}{Faloutsos, C.}, \bibinfo{year}{2005}.
\newblock \bibinfo{title}{Graphs over time: densification laws, shrinking diameters and possible explanations}, in: \bibinfo{booktitle}{Proceedings of the eleventh ACM SIGKDD international conference on Knowledge discovery in data mining}, pp. \bibinfo{pages}{177--187}.
\bibitem[{Liu et~al.(2024)Liu, Devunuri, Dong, Lehe and Gayah}]{liu2024impact}
\bibinfo{author}{Liu, H.}, \bibinfo{author}{Devunuri, S.}, \bibinfo{author}{Dong, X.}, \bibinfo{author}{Lehe, L.}, \bibinfo{author}{Gayah, V.}, \bibinfo{year}{2024}.
\newblock \bibinfo{title}{Impact of competition on the scale effects in ridesplitting: A case study of manhattan}.
\newblock \bibinfo{journal}{103rd Annual Meeting of the Transportation Research Board} \URLprefix \url{https://par.nsf.gov/biblio/10499938}.
\bibitem[{Liu et~al.(2023)Liu, Devunuri, Lehe and Gayah}]{liu2023scale}
\bibinfo{author}{Liu, H.}, \bibinfo{author}{Devunuri, S.}, \bibinfo{author}{Lehe, L.}, \bibinfo{author}{Gayah, V.V.}, \bibinfo{year}{2023}.
\newblock \bibinfo{title}{Scale effects in ridesplitting: A case study of the city of chicago}.
\newblock \bibinfo{journal}{Transportation Research Part A: Policy and Practice} \bibinfo{volume}{173}, \bibinfo{pages}{103690}.
\bibitem[{Masoud and Jayakrishnan(2017)}]{masoud2017decomposition}
\bibinfo{author}{Masoud, N.}, \bibinfo{author}{Jayakrishnan, R.}, \bibinfo{year}{2017}.
\newblock \bibinfo{title}{A decomposition algorithm to solve the multi-hop peer-to-peer ride-matching problem}.
\newblock \bibinfo{journal}{Transportation Research Part B: Methodological} \bibinfo{volume}{99}, \bibinfo{pages}{1--29}.
\bibitem[{Meshkani and Farooq(2022)}]{meshkani2022generalized}
\bibinfo{author}{Meshkani, S.M.}, \bibinfo{author}{Farooq, B.}, \bibinfo{year}{2022}.
\newblock \bibinfo{title}{A generalized ride-matching approach for sustainable shared mobility}.
\newblock \bibinfo{journal}{Sustainable Cities and Society} \bibinfo{volume}{76}, \bibinfo{pages}{103383}.
\bibitem[{{NYC Taxi and Limousine Commission}(2024)}]{tlc}
\bibinfo{author}{{NYC Taxi and Limousine Commission}}, \bibinfo{year}{2024}.
\newblock \bibinfo{title}{Nyc tlc trip record data}.
\newblock \bibinfo{howpublished}{https://www.nyc.gov/site/tlc/about/tlc-trip-record-data.page}.
\bibitem[{Pandey et~al.(2025)Pandey, Liu, Dong, Lehe and Gayah}]{pandey2025decline}
\bibinfo{author}{Pandey, A.}, \bibinfo{author}{Liu, H.}, \bibinfo{author}{Dong, X.}, \bibinfo{author}{Lehe, L.}, \bibinfo{author}{Gayah, V.V.}, \bibinfo{year}{2025}.
\newblock \bibinfo{title}{Decline of ride-splitting: A case study of new york city}.
\newblock \bibinfo{journal}{National Academies of Engineering} \URLprefix \url{https://par.nsf.gov/biblio/10584875}.
\bibitem[{Qin et~al.(2022)Qin, Zhu and Ye}]{qin2022reinforcement}
\bibinfo{author}{Qin, Z.T.}, \bibinfo{author}{Zhu, H.}, \bibinfo{author}{Ye, J.}, \bibinfo{year}{2022}.
\newblock \bibinfo{title}{Reinforcement learning for ridesharing: An extended survey}.
\newblock \bibinfo{journal}{Transportation Research Part C: Emerging Technologies} \bibinfo{volume}{144}, \bibinfo{pages}{103852}.
\bibitem[{Santi et~al.(2014)Santi, Resta, Szell, Sobolevsky, Strogatz and Ratti}]{santi2014quantifying}
\bibinfo{author}{Santi, P.}, \bibinfo{author}{Resta, G.}, \bibinfo{author}{Szell, M.}, \bibinfo{author}{Sobolevsky, S.}, \bibinfo{author}{Strogatz, S.H.}, \bibinfo{author}{Ratti, C.}, \bibinfo{year}{2014}.
\newblock \bibinfo{title}{Quantifying the benefits of vehicle pooling with shareability networks}.
\newblock \bibinfo{journal}{Proceedings of the National Academy of Sciences} \bibinfo{volume}{111}, \bibinfo{pages}{13290--13294}.
\bibitem[{S{\'e}journ{\'e} et~al.(2018)S{\'e}journ{\'e}, Samaranayake and Banerjee}]{sejourne2018price}
\bibinfo{author}{S{\'e}journ{\'e}, T.}, \bibinfo{author}{Samaranayake, S.}, \bibinfo{author}{Banerjee, S.}, \bibinfo{year}{2018}.
\newblock \bibinfo{title}{The price of fragmentation in mobility-on-demand services}.
\newblock \bibinfo{journal}{Proceedings of the ACM on Measurement and Analysis of Computing Systems} \bibinfo{volume}{2}, \bibinfo{pages}{1--26}.
\bibitem[{Shaheen and Cohen(2019)}]{shaheen2019shared}
\bibinfo{author}{Shaheen, S.}, \bibinfo{author}{Cohen, A.}, \bibinfo{year}{2019}.
\newblock \bibinfo{title}{Shared ride services in north america: definitions, impacts, and the future of pooling}.
\newblock \bibinfo{journal}{Transport reviews} \bibinfo{volume}{39}, \bibinfo{pages}{427--442}.
\bibitem[{Shapley et~al.(1953)}]{shapley1953value}
\bibinfo{author}{Shapley, L.S.}, et~al., \bibinfo{year}{1953}.
\newblock \bibinfo{title}{A value for n-person games}.
\newblock \bibinfo{journal}{Princeton University Press Princeton} .
\bibitem[{Shulika et~al.(2024)Shulika, Bujak, Ghasemi and Kucharski}]{shulika2024spatiotemporal}
\bibinfo{author}{Shulika, O.}, \bibinfo{author}{Bujak, M.}, \bibinfo{author}{Ghasemi, F.}, \bibinfo{author}{Kucharski, R.}, \bibinfo{year}{2024}.
\newblock \bibinfo{title}{Spatiotemporal variability of ride-pooling potential--half a year new york city experiment}.
\newblock \bibinfo{journal}{Journal of Transport Geography} \bibinfo{volume}{114}, \bibinfo{pages}{103767}.
\bibitem[{Stigler(1958)}]{stigler1958economies}
\bibinfo{author}{Stigler, G.J.}, \bibinfo{year}{1958}.
\newblock \bibinfo{title}{The economies of scale}.
\newblock \bibinfo{journal}{The Journal of Law and Economics} \bibinfo{volume}{1}, \bibinfo{pages}{54--71}.
\bibitem[{Sundt et~al.(2021)Sundt, Luo, Vincent, Shahabi and Yin}]{sundt2021heuristics}
\bibinfo{author}{Sundt, A.}, \bibinfo{author}{Luo, Q.}, \bibinfo{author}{Vincent, J.}, \bibinfo{author}{Shahabi, M.}, \bibinfo{author}{Yin, Y.}, \bibinfo{year}{2021}.
\newblock \bibinfo{title}{Heuristics for customer-focused ride-pooling assignment}.
\newblock \bibinfo{journal}{arXiv preprint arXiv:2107.11318} .
\bibitem[{Tachet et~al.(2017)Tachet, Sagarra, Santi, Resta, Szell, Strogatz and Ratti}]{tachet2017scaling}
\bibinfo{author}{Tachet, R.}, \bibinfo{author}{Sagarra, O.}, \bibinfo{author}{Santi, P.}, \bibinfo{author}{Resta, G.}, \bibinfo{author}{Szell, M.}, \bibinfo{author}{Strogatz, S.H.}, \bibinfo{author}{Ratti, C.}, \bibinfo{year}{2017}.
\newblock \bibinfo{title}{Scaling law of urban ride sharing}.
\newblock \bibinfo{journal}{Scientific reports} \bibinfo{volume}{7}, \bibinfo{pages}{1--6}.
\bibitem[{Tafreshian et~al.(2020)Tafreshian, Masoud and Yin}]{tafreshian2020frontiers}
\bibinfo{author}{Tafreshian, A.}, \bibinfo{author}{Masoud, N.}, \bibinfo{author}{Yin, Y.}, \bibinfo{year}{2020}.
\newblock \bibinfo{title}{Frontiers in service science: Ride matching for peer-to-peer ride sharing: A review and future directions}.
\newblock \bibinfo{journal}{Service Science} \bibinfo{volume}{12}, \bibinfo{pages}{44--60}.
\bibitem[{Vignon et~al.(2023)Vignon, Yin and Ke}]{vignon2023regulating}
\bibinfo{author}{Vignon, D.}, \bibinfo{author}{Yin, Y.}, \bibinfo{author}{Ke, J.}, \bibinfo{year}{2023}.
\newblock \bibinfo{title}{Regulating the ride-hailing market in the age of uberization}.
\newblock \bibinfo{journal}{Transportation research part E: logistics and transportation review} \bibinfo{volume}{169}, \bibinfo{pages}{102969}.
\bibitem[{Wang et~al.(2023)Wang, Zhao, Zhang, Guo and Zhao}]{wang2023quantifying}
\bibinfo{author}{Wang, X.}, \bibinfo{author}{Zhao, Z.}, \bibinfo{author}{Zhang, H.}, \bibinfo{author}{Guo, X.}, \bibinfo{author}{Zhao, J.}, \bibinfo{year}{2023}.
\newblock \bibinfo{title}{Quantifying the uneven efficiency benefits of ridesharing market integration}.
\newblock \bibinfo{journal}{arXiv preprint arXiv:2303.13520} .
\bibitem[{Wang et~al.(2022)Wang, Tong, Zhou, Ren, Xu, Wu and Lv}]{wang2022fed}
\bibinfo{author}{Wang, Y.}, \bibinfo{author}{Tong, Y.}, \bibinfo{author}{Zhou, Z.}, \bibinfo{author}{Ren, Z.}, \bibinfo{author}{Xu, Y.}, \bibinfo{author}{Wu, G.}, \bibinfo{author}{Lv, W.}, \bibinfo{year}{2022}.
\newblock \bibinfo{title}{Fed-ltd: Towards cross-platform ride hailing via federated learning to dispatch}, in: \bibinfo{booktitle}{Proceedings of the 28th ACM SIGKDD Conference on Knowledge Discovery and Data Mining}, pp. \bibinfo{pages}{4079--4089}.
\bibitem[{Xu et~al.(2022)Xu, AMC~Vignon, Yin and Ye}]{xu2022empirical}
\bibinfo{author}{Xu, Z.}, \bibinfo{author}{AMC~Vignon, D.}, \bibinfo{author}{Yin, Y.}, \bibinfo{author}{Ye, J.}, \bibinfo{year}{2022}.
\newblock \bibinfo{title}{An empirical study of the labor supply of ride-sourcing drivers}.
\newblock \bibinfo{journal}{Transportation Letters} \bibinfo{volume}{14}, \bibinfo{pages}{352--355}.
\bibitem[{Ying et~al.(2021)Ying, Dong, Li and Tian}]{ying2021auto}
\bibinfo{author}{Ying, J.}, \bibinfo{author}{Dong, X.}, \bibinfo{author}{Li, B.}, \bibinfo{author}{Tian, Z.}, \bibinfo{year}{2021}.
\newblock \bibinfo{title}{Auto-regressive model with exogenous input (arx) based traffic flow prediction}, in: \bibinfo{booktitle}{CICTP 2021}, pp. \bibinfo{pages}{295--304}.
\bibitem[{Zhong et~al.(2020)Zhong, Zhang, Nie and Xu}]{zhong2020dynamic}
\bibinfo{author}{Zhong, L.}, \bibinfo{author}{Zhang, K.}, \bibinfo{author}{Nie, Y.M.}, \bibinfo{author}{Xu, J.}, \bibinfo{year}{2020}.
\newblock \bibinfo{title}{Dynamic carpool in morning commute: Role of high-occupancy-vehicle (hov) and high-occupancy-toll (hot) lanes}.
\newblock \bibinfo{journal}{Transportation Research Part B: Methodological} \bibinfo{volume}{135}, \bibinfo{pages}{98--119}.

\end{thebibliography}

\end{document}